\documentclass{aa}  

\usepackage{graphicx}
\usepackage{txfonts}
\begin{document}

   \title{Panchromatic calibration of Ca II triplet luminosity dependence}

   \author{B. Dias
          \inst{1}
          \and
          M. C. Parisi\inst{2,3}
          }

\institute{
   	Instituto de Alta Investigaci\'on, Universidad de Tarapac\'a, Casilla 7D, Arica, Chile\\  
              \email{bdiasm@academicos.uta.cl}
         \and
   	Observatorio Astron\'omico, Universidad Nacional de C\'ordoba, Laprida 854, X5000BGR, C\'ordoba, Argentina.\\  
              \email{cparisi@unc.edu.ar}
         \and
         Instituto de Astronom{\'\i}a Te\'orica y Experimental (CONICET-UNC), Laprida 854, X5000BGR, C\'ordoba, Argentina.
             }

   \date{Received ...; accepted ...}

 
  \abstract
   {The line strength of the near infrared Ca II triplet (CaT) lines are a proxy to measure metallicity from integrated and individual stellar spectra of bright red giant stars. In the latter case it is a mandatory step to remove the magnitude (proxy for gravity, temperature and luminosity) dependence from the equivalent width of the lines before converting them into metallicities. The working empirical procedure used for decades is to use the relative magnitude with respect to the horizontal branch level or red clump, with the advantage of being independent from distance and extinction.}
   {The $V$ filter is broadly adopted as the reference magnitude, although a few works have used different filters ($I$ and $Ks$, for example). In this work we investigate the dependence of the CaT calibration using {\it $griz$} filters from the Dark Energy Camera (DECam) and the Gemini Multi-Object Spectrograph (GMOS), {\it G} from Gaia, {\it BVI} filters from the Magellanic Clouds photometric survey (MCPS), {\it $YJKs$} filters from Visible and Infrared Survey Telescope for Astronomy (VISTA) InfraRed CAMera (VIRCAM). We use as a reference the FOcal Reducer and low dispersion Spectrograph 2 (FORS2) $V$ filter used in the original analysis of the sample.
   }
   {Red giant stars from clusters with known metallicity and available CaT equivalent widths are used as reference. Public photometric catalogues are taken from the Survey of the MAgellanic Stellar History (SMASH) second data release, VISTA survey of the Magellanic Clouds system (VMC), Gaia, MCPS surveys plus VIsible Soar photometry of star Clusters in tApi'i and Coxi HuguA (VISCACHA)-GMOS data, for a selection of Small Magellanic Cloud clusters.
   The slopes are fitted using two and three lines to be applicable to most of the metallicity scales.}
   {The magnitude dependence of the CaT equivalent widths is well described by a linear relation using any filter analysed in this work. The slope increase with wavelength of the filters. The zero point (a.k.a. reduced equivalent width), that is the metallicity indicator, remains the same.}
   {If the same line profile function is used with the same bandpasses and continuum regions, and the total equivalent width (EW) comes from the same number of lines (2 or 3), then the reduced EW is the same regardless the filter used. Therefore, any filter can be used to convert the CaT equivalent widths into metallicity for a given CaT calibration.}


   \maketitle
\section{Introduction}

The near infrared Ca II triplet (CaT) lines at 8498, 8542, 8662 ${\rm \AA}$ are strong lines that have been used to estimate metallicities in late-type bright stars as early as 1960-1970 \citep[e.g.][]{spinrad+69}. The method currently used to convert the CaT line strength into metallicity was first proposed by \cite{armandroff+88} and \cite[][AD91]{armandroff+91} for integrated spectra and individual red giant stars of globular clusters, and \cite{olszewski+91} for red giant stars of Large Magellanic Cloud clusters, although alternative techniques were used before that \citep[e.g.][]{diaz+89}. The technique has been fine-tuned and discussed in detail in many subsequent works since then. For instance, the effect of age and population \citep[e.g.][]{cole+04,pont+04,vasquez+18}, the effect of [Ca/Fe] ratio \citep[e.g.][]{battaglia+08,dacosta16}, metal-rich stars \citep[e.g.][]{carrera+07,vasquez+15}, metal-poor stars \citep[e.g.][]{starkenburg+10,carrera+13}, reliability of integrated CaT \citep[e.g.][]{usher+19}, inclusion of stars fainter than the horizontal branch level \citep[e.g.][]{husser+20}.

The equivalent width (EW) of an atomic line is sensitive to the surface gravity log($g$), effective temperature T$_{\rm eff}$, overall metallicity [Fe/H], micro-turbulence velocity, oscillator strength and finally the specific element abundance \citep[e.g.][]{gray08, barbuy+08}. It has been demonstrated that in the case of the saturated\footnote{The EW of the CaT lines is more sensitive to the wings than to the core of the lines \citep[e.g.][]{erdelyi-mendes+91}} CaT lines, they are more sensitive to [Fe/H] than to [Ca/Fe] itself (e.g. \citealp{battaglia+08,dacosta16}, but see also \citealp{bosler+07}; \citealp{starkenburg+10}). Micro-turbulence velocity mostly affects metal-poor stars closer to the tip of the red giant branch \citep[RGB,][]{starkenburg+10}, which means this is a negligible effect for stars in the Magellanic Clouds, specially those closer to the red clump (RC). Therefore, the CaT can be used as a proxy for [Fe/H] with confidence, given that the dependence on the other parameters can be removed.

Luminosity of RGB stars correlate well with log($g$) and T$_{\rm eff}$ and can be used as proxy to remove gravity and temperature effects on the CaT EW, leaving only the dependence on [Fe/H], as discussed by AD91. They argued in favour of using a relative magnitude instead of absolute magnitude to avoid the dependence on the distance and extinction of each star. They tested and calibrated the technique using the commonly used $V$ band magnitude relative to the magnitude of the horizontal branch (HB). This filter has been broadly used to calibrate CaT metallicities, although some works have applied other filters (e.g. $I$ by \citealp{carrera+07}; K$_s$ by \citealp{mauro+14} and \citealp{carrera+13}; $F606W$, $F555W$, $F625W$ by \citealp{husser+20}; $G$ by \citealp{simpson20}). Not always there is published photometry in the appropriate filter to calibrate the observed CaT. For instance, \cite{olszewski+91} solved this difficulty by creating a magnitude from their spectra ($m_{8600}$). 
Another reason for employing different filters is the environment, for example, infrared filters are more suitable for CaT of Milky Way bulge stars.

Notwithstanding the apparent flexibility of using different filters, each work had to use calibration clusters to check their conversion of CaT EW into metallicities. In this paper we investigate whether it is possible to follow exactly the recipes of a given CaT calibration, change only the filter used and find the same final metallicities without the need of calibrating again using reference stars. If so, any CaT study would not have the necessity of calibrating a new scale for each analysis, and could choose a CaT calibration among the plethora of works, and apply it with any available photometry.

In particular, we take the sample from \citet[][P09]{parisi+09} and \citet[][P15]{parisi+15} who used the CaT calibration and recipes of \citet[][C04]{cole+04} and $V$ magnitudes to derive metallicities for Small Magellanic Cloud (SMC) clusters. Specifically, C04 use a combination of Gaussian+Lorentzian to fit the CaII lines, the definition $\sum EW = EW_{8498} + EW_{8542} + EW_{8662}$ and bandpasses defined by \cite{armandroff+88}.
The same CaT calibration and recipes of C04 are kept while changing the filters to investigate the impact on the final metallicities. It is worth noting that C04 analysed clusters with ages between 2.5 and 14~Gyr and found no significant age effect on their metallicity calibration. \cite{pont+04} only found significant age effect on the CaT metallicity calibration if stars with ages below 1~Gyr are compared to stars with 12.6~Gyr. Therefore, it is safe to say that the CaT technique can be applied to stars with ages above 1-2~Gyr, which is the case for the clusters analysed in this paper (see Sect. \ref{sec:sample}).

This analysis will have immediate use on the ongoing spectroscopic follow-up of the VIsible Soar photometry of star Clusters in tApi'i and Coxi HuguA (VISCACHA) survey \citep{maia+19} that uses GMOS/Gemini to observe star clusters in the Magellanic Clouds\footnote{\url{ http://www.astro.iag.usp.br/~viscacha/}} (PI: Dias). This instrument offers $gri$ filters for the pre-images, and they have not been used to calibrate CaT metallicities so far, and the $VI$ magnitudes are available only for a fraction of the stars from the VISCACHA survey and Magellanic Clouds photometric survey (MCPS); therefore a new calibration using $gri$ filters is required. Moreover, the Magellanic Clouds have been gaining interest and being observed by many photometric surveys (e.g. s VISTA survey of the Magellanic Clouds system, VMC, \citealp{cioni+11}; Survey of the MAgellanic Stellar History, SMASH, \citealp{nidever+17}; VISCACHA, \citealp{maia+19} and others). The respective spectroscopic follow-up campaigns have started or are about to start with the 4-metre Multi-Object Spectrograph Telescope (4MOST), Apache Point Observatory Galactic Evolution Experiment (APOGEE) and Gemini Multi-Object Spectrograph (GMOS), respectively. 4MOST will also perform a follow-up of VMC that includes the CaT, therefore they would benefit from a CaT calibration with their $YJK_s$ filters from VMC\footnote{\url{https://www.eso.org/sci/facilities/paranal/instruments/vircam/inst.html}}. Also, SMASH\footnote{\url{http://www.ctio.noao.edu/noao/node/13140}} uses the same filters as GMOS\footnote{\url{http://www.gemini.edu/sciops/instruments/gmos/imaging/filters}}, and therefore their photometry could be applied to other CaT studies using the analysis of this paper.

The CaT technique is a powerful tool to determine metallicities for individual stars in distant galaxies as the CaII lines are very strong and their wavelength matches the peak flux of typical RGB stellar spectrum (see Fig. \ref{fig:filters}). For example, \cite{bosler+07} studied stars from LeoI and LeoII located at 273 and 204~kpc away; \cite{gilbert+06} and \cite{koch+08} studied stars in M\,31 located at about 784~kpc away. This technique will continue to be very useful when the European Southern Observatory (ESO) Extremely Large Telescope (ELT) is available to observe CaT in resolved stars beyond the Local Group \citep[see][]{battaglia11}.

Different CaT calibrations make use of the sum of the EW of the three CaT lines or only the two strongest lines. We also present here the impact that the different choices of filters have in the final metallicities if the calibration uses two or three lines. Other choices that differ from calibration to calibration are not analysed in this paper, as they have been discussed in many previous works, such as: are the function fitted to the line profile (as discussed e.g. by \citealp{saviane+12,husser+20}), the line and continuum bandpasses (as discussed e.g. by \citealp{carrera+07}), continuum level and normalisation (as discussed e.g. by \citealp{vasquez+15}).

This paper is organised as follows: the analysis of isochrones to make the predictions discussed in this work is discussed in Sect. 2. The sample selection and data are described in Sect. 3. The core analysis of the paper is presented in Sect. 4, where the fitting results are compared to the theoretical predictions from Sect.2. Finally, we check the systematic errors that the filter choice makes in the final metallicity in Sect. 5. Summary and conclusions can be found in Sect. 6.

\section{RGB relative slope for different filters}
\label{sec:isoc}

The motivation of this work is to isolate only one aspect of the CaT calibration, i.e., the luminosity indicator used as a proxy to erase the log($g$) and T$_{\rm eff}$ dependence of the CaT EW. Therefore, it would be expected that the variation on the CaT calibration comes from the relative slope of the RGB depending on the filter choice. In this first section, we use PAdova and TRieste Stellar Evolution Code (PARSEC) isochrones `CMD 3.3 version'\footnote{\url{http://stev.oapd.inaf.it/cgi-bin/cmd}} \citep{bressan+12} to derive this dependence. We chose this set of isochrones as they provide all sets of filters we analyse in this work, thus producing a homogeneous output analysis.
Isochrones were downloaded for $8.0\leq$ log($age$) $\leq10.1$ in steps of 0.1~dex, and $-1.5\leq$ [M/H] $\leq-0.1$ in steps of 0.1 dex that represent the overall SMC cluster population. All sets of filters were chosen: $UBVRIJHKLM$, $VISTA ZYJHKs$, $DECAM$, and $Gaia$ \citep{gaia2016}.

All isochrones within this age and metallicity range present a RC, therefore it is straightforward to determine their HB level, assumed here as the RC level for simplicity. This is the reference for the relative magnitude used in the CaT calibration. Only the RGB region is isolated, and the magnitudes relative to the RC level are calculated for all filters.
As a first approximation, we fit a linear relation 

\begin{equation}
(V-V_{HB}) = a + b\cdot(m-m_{HB})   
\label{eq:magrel}
\end{equation}

\noindent for RGB stars and all filters $m$. The reference on the filter $V$ is chosen because this is the most popular filter used in CaT calibrations, and therefore the relations derived here can be used to scale any calibration using $V$ to other filters. 

\begin{figure}[!htb]
    \centering
    \includegraphics[width=\columnwidth]{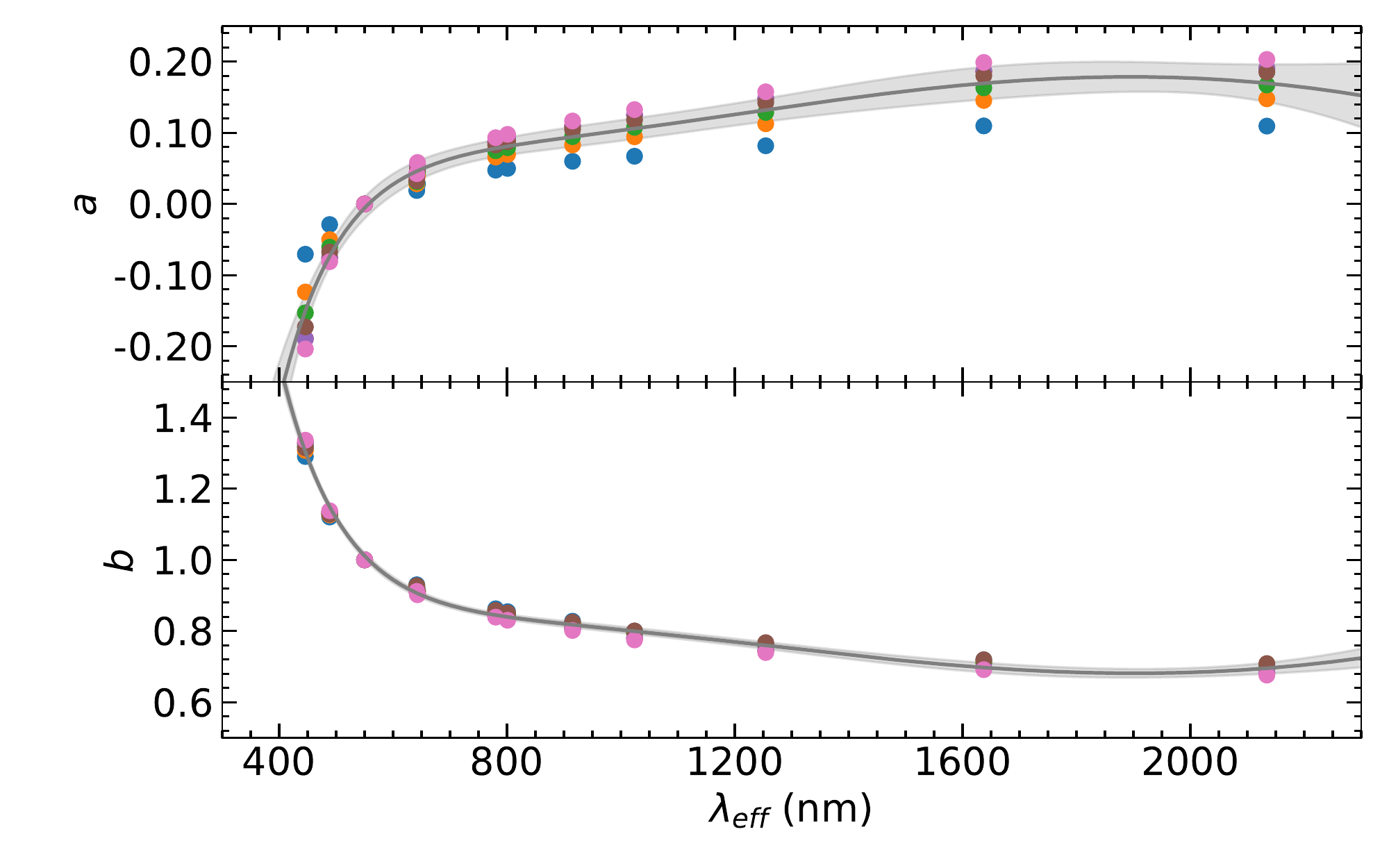}
    \caption{Coefficients $a,b$ from Eq. \ref{eq:magrel} as a function of the effective wavelength for each filter $m$, where $m = B,g,V,G,r,i,I,z,Y,J.H,Ks$. We note that the filter $H$ was included here to constrain better the fit, even though it is not present in the analysis in the rest of the paper. The fit was performed for isochrones with combinations of age and metallicity for the sample clusters from Table \ref{tab:sample}, which are represented by different colours, mostly overlapping points. The line is a polynomial fit to all points and the shaded area represents $\pm3\sigma$ uncertainties.
    }
    \label{fig:isoctypical}
\end{figure}

The coefficients $a,b$ from Eq. \ref{eq:magrel} are shown in Fig. \ref{fig:isoctypical} for all filters $m$, where $m = B,g,V,G,r,i,I,z,Y,J,H,Ks$. We show that these coefficients are little sensitive to variations in age and metallicity in the range $[1.5, 5.0]$ Gyr and $[-1.2, -0.6]$ dex, and their dependence with wavelength is well represented by

\begin{equation}
    \begin{aligned}
    a = -9.8(\pm 2.1)\cdot x^4 + 120(\pm 25)\cdot x^3 - 545(\pm 111)\cdot x^2 \\
    + 1102(\pm 220)\cdot x - 836(\pm163){\rm , and}
    \end{aligned}
    \label{eq:a-poly}
\end{equation}

\begin{equation}
    \begin{aligned}
    b = 18.3(\pm 1.2)\cdot x^4 - 224(\pm 15)\cdot x^3 + 1023(\pm 65)\cdot x^2 \\
    - 2078(\pm 129)\cdot x + 1583(\pm96){\rm ,}
    \end{aligned}
    \label{eq:b-poly}
\end{equation}

\noindent where $x = \log_{10}(\lambda_{eff}/nm)$.

The reduced EW $W'$ is the zero point of the equation below, that is essentially a linear function of $\sum EW$ against $(m-m_{HB})$:

\begin{equation}
\sum EW = W'_m - \beta_m\cdot(m-m_{HB}).
\label{eq:REWV}
\end{equation}

\noindent If we write Eq. \ref{eq:REWV} for $m=V$ and replace $(V-V_{HB})$ by the right side of Eq. \ref{eq:magrel}, the coefficients for a given filter $m$ are $W'_m = W'_V - \beta_V\cdot a$ and $\beta_m = \beta_V\cdot b$.

Assuming $\beta_V =  0.73\ \AA/mag$ from C04, the variation in $W'_m$ around $W'_V$ is about $\pm 0.15\ \AA/mag$, which is of the order of the uncertainties in $W'_m$. The blue points in Fig. \ref{fig:isoctypical} top panel refer to an age of 1.5~Gyr (and [Fe/H] = -0.6) and stand out from the relation, while the orange points for 2.0~Gyr have similar zero points $a$ as for the other ages. Nevertheless, the residuals of the blue points w.r.t. Eq. \ref{eq:a-poly} of $\lesssim 0.05$ mean a variation in $W'_m$ around $W'_V$ of about $\pm 0.04\ \AA/mag$, which is much smaller than the errors in $W'_m$. Therefore, it is expected that $W'_m$ does not change with the filter, within the typical uncertainties. 
On the other hand, the variation of $\beta_m$ from about $1.0$ to $0.5$ for $\beta_V =  0.73\ \AA/mag$ cannot be explained by uncertainties, and it is a real detectable difference.

We collect observed photometry for a selection of clusters on all filters in the next section to check whether the family of curves for $\beta_m$ represent well the observations and to confirm whether $W'_m$ is constant within uncertainties for any filter choice.

\section{Sample selection and data}
\label{sec:sample}

\begin{figure}[!htb]
\centering
\includegraphics[width=\columnwidth]{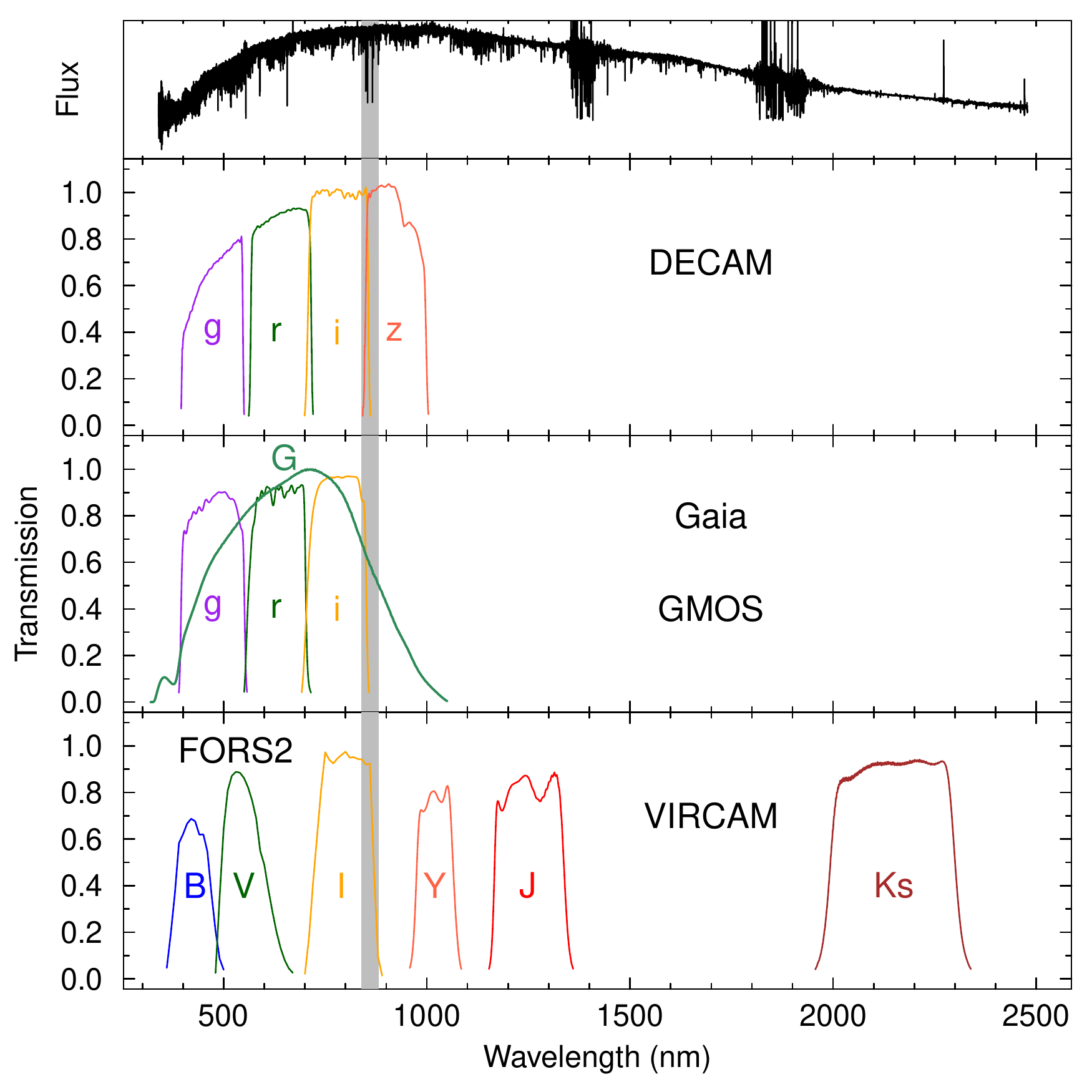}
   \caption{
   Transmission curves of the filters used in this paper from DECam, GMOS, Gaia, FORS2\protect\footnotemark, and VIRCAM. The spectrum of the typical RGB star 2MASS J16232694-2631313 from the XSHOOTER spectral library \citep{gonneau+20} with (T$_{\rm eff}$, log($g$), [Fe/H]) $\approx$ (4500K, 1.6, -1.0) \citep{arentsen+19} is displayed as a reference. The CaT region is highlighted in grey and matches the peak flux of the spectrum.}
      \label{fig:filters}
\end{figure}

The sample selection is based on the initial sample of 29 clusters from P09 and P15 together. A subsample was then selected based on whether there is available photometry in all filters from SMASH data release 2 (DR2), VMC DR5, MCPS surveys to be analysed together the photometry from the original analysis with the FOcal Reducer and low dispersion Spectrograph 2 for the Very Large Telescope (FORS2/VLT) and from the all-sky Gaia survey. We also rejected four clusters that were younger than $\sim$1.5 Gyr, below which the RC level is higher or absent and the standard CaT calibration is no longer valid. The third criterion of sample selection is related to our choice to focus only on the variation of filters in the calibration, leaving other variables to other works. Theoretically, the slope $\beta$ in the relation $\sum EW$ vs. $m-m_{HB}$ varies with metallicity \citep[e.g.][]{jorgensen+92,starkenburg+10,carrera+13,husser+20}. This relation seems to be detectable only for extreme metallicities. Besides, there seems to be an intrinsic dispersion in $\beta$ from cluster to cluster not correlated to metallicity. For example, for the clusters used as calibrators by C04, $\beta$ varies from $\sim0.4$ to $\sim0.9\ \AA/mag$ with the $V$ filter. C04 reported a relatively large intrinsic scatter in the $\sum EW$ for a given single-metallicity cluster, also found by \cite{rutledge+97a}. We find a similar r.m.s. scatter, reported in Table \ref{tab:W}.
The dispersion in $\beta$ will not be discussed in this paper, therefore we further select clusters from P09, P15 that have $0.4 \lesssim \beta_V \lesssim 0.9\ \AA/mag$ to end up with a {\it bona-fide} sample that can reproduce $\beta_V$ from C04, and find the respective $\beta_m$ for all other filters displayed in Fig. \ref{fig:filters}. The final sample of seven SMC clusters is presented in Table \ref{tab:sample}.

\begin{table}[!htb]
\caption{SMC cluster sample sorted by age. Metallicities come from P09,P15 and the references for ages are specified for each cluster.
}             
\label{tab:sample}      
\centering                          
\begin{tabular}{l c c c}        
\hline\hline                 
Cluster & [Fe/H] & Age (Gyr) & Ref.  \\    
\hline                        
\object{Kron\,6}      & -0.63$\pm$0.02 & 1.6 & P05 \\
\object{HW\,67}       & -0.72$\pm$0.04 & 1.9 & P17 \\
\object{HW\,47}       & -0.92$\pm$0.04 & 3.3 & P14 \\
\object{HW\,40}       & -0.78$\pm$0.05 & 2.5 & D14 \\
\object{Lindsay\,17}  & -0.84$\pm$0.03 & 4.4 & P14 \\
\object{Kron\,9}      & -1.12$\pm$0.05 & 4.7 & G10  \\
\object{Lindsay\,19}  & -0.85$\pm$0.03$^a$ & 4.8& P14  \\
\hline                                   
\end{tabular}
\tablefoot{P14: \cite{parisi+14}; G10: \cite{glatt+10}; P05: \cite{piatti+05}; P17: \cite{perren+17}; D14: \cite{dias+14}; (a) Only 6 out of 7 stars from P09 were found in all catalogues, therefore the reference [Fe/H] for Lindsay\,19 is updated accordingly for consistency, although with no significant change.
}
\end{table}

The spectroscopic data observed with FORS2/VLT was not re-analysed here, the cluster membership and equivalent width for the three CaT lines for all stars are taken directly from P09 and P15 to be fully consistent regarding the spectroscopic analysis and focus only on the filter choice. We also use the $V$ magnitudes from their original work. We derived again the HB level of the P09, P15 original $V$ photometry using our method described in the next subsection and the values are compatible, therefore we keep $V-V_{HB}$ directly from P09, P15.

\footnotetext{\url{http://www.eso.org/sci/facilities/paranal/instruments/fors/inst/Filters/curves.html}}

The photometry from the SMASH DR2\footnote{\url{https://datalab.noao.edu/smash/smash.php}} \citep{nidever+17} was obtained to analyse the Dark Energy Camera (DECam) $griz$ filters.
An original motivation of this work was to derive a CaT calibration valid for the GMOS $gri$ photometry from the pre-images of the spectroscopic follow-up of the VISCACHA survey, but there is no cluster in common with P09, P15. In Sect. \ref{sec:gmosdecam} we show that $m-m_{HB}$ is equivalent for DECam and GMOS for the $gri$ filters. Therefore, all the analysis performed using DECam $gri$ filters is automatically valid for the GMOS filters.
The photometry from the MCPS\footnote{\url{https://www.as.arizona.edu/~dennis/mcsurvey}} \citep{zaritsky+02} and Gaia DR2\footnote{\url{https://www.cosmos.esa.int/web/gaia/data-release-2}} \citep{gaia2018} surveys was obtained to analyse the filters $BVI$ and $G$.

The near-infrared photometry from VMC DR5\footnote{\url{https://www.eso.org/sci/publications/announcements/sciann17232.html}} \citep{cioni+11} was obtained in the filters $YJKs$. At the moment when this paper was submitted, there was only multi-aperture photometry available with non-calibrated fluxes. Nevertheless, aperture correction and zero points make no difference in our analysis because we use relative magnitudes and these terms cancel out. Therefore, we chose a single aperture flux ($aper3$, recommended in the VMC DR5 documentation) and simply calculated $-2.5\cdot\log_{10}\ (aper3)$. We downloaded a single tile catalogue per filter, with ESO grade $A$ or $B$, meaning the observations met the weather constraints. We found this procedure to be enough to produce reasonable colour-magnitude diagrams (CMDs) of the RGB stars down to the RC level.

\subsection{The horizontal branch level}

As defined by AD91, $(m-m_{HB})$ is a proxy for gravity and has the double advantage of being model and photometric calibration independent. As a consequence, the definition of the HB magnitude level is a crucial step. 

\begin{figure}[!htb]
   \centering
	\includegraphics[width=0.65\columnwidth]{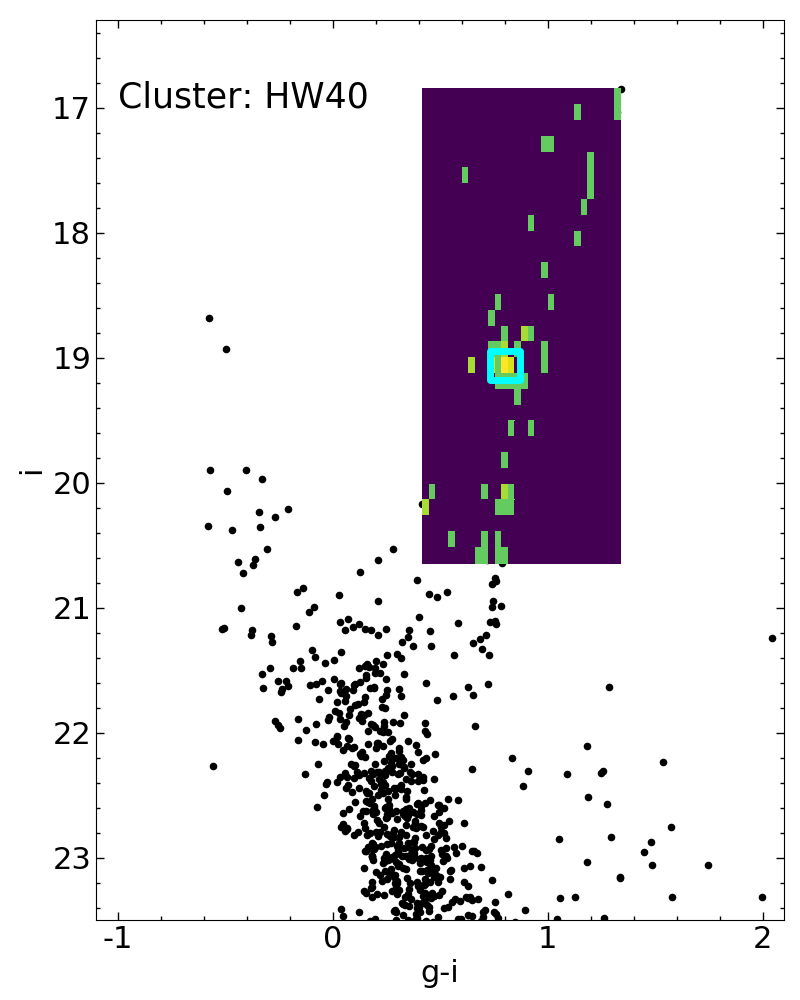}
      \caption{Example of RC fit for HW\,40 from the calibrated SMASH DR2 $gi$ photometry. Black dots are the stars within the cluster radius of $4.5'$ \citep{bica+20}. The selected RGB region is coloured by the rectangle where the dots are replaced by the 2d histogram bins. The clear over-density matches the fitted RC position indicated by the cyan square. 
      }
         \label{fig:RC}
   \end{figure}

The Magellanic Cloud clusters analysed here are relatively metal-rich, which means that their HB is essentially a RC. Therefore, we assume the HB level as the magnitude of the RC for simplicity. 
We make an initial selection by eye around the RGB zone containing the RC on the CMD composed of stars within the cluster size as provided by \cite{bica+20}. We then used a python script written by us to find the position of the RC. Because the RC is clearly identified by eye, it was enough to generate a 2D histogram (with bin size big enough to be able to visually identify the RC, which was typically three times the error in colour) and find the position of its peak (see example in Fig. \ref{fig:RC}).
In some cases, the RC was underpopulated and therefore some interaction was required to fine tune the bin size and cluster radius until converge to a visually satisfactory position of the RC, consistent with the other clusters for all filters. The final ``HB magnitude'' and respective error is the average and standard deviation of the magnitude of all stars within the four bins around the maximum. The same procedure is done for all filters from Fig. \ref{fig:filters} and clusters listed in Table \ref{tab:sample}.

%
\subsection{GMOS/Gemini vs. DECam/Blanco filters}
\label{sec:gmosdecam}

The GMOS data was collected within the projects GS-2017B-Q-19 (PI: Kerber, 5 clusters with $ri$ filters), GS-2018B-Q-208, GS-2018B-Q-302, GS-2019B-Q-303 (PI: Dias for the last three, 21 clusters with $gr$ filters). The match with SMASH DR2 photometry resulted in 4 clusters with $ri$ filters and 19 clusters with $gr$ filters.

The transmission curves of the $gri$ filters from GMOS and DECam have similar wavelength range,
but with slightly different transmission curves. The differences should be cancelled when comparing the differential magnitudes with respect to the HB (RC) level, which is the information required for the CaT calibration. We check if this is true by direct comparing $(m-m_{\rm HB})_{\rm GMOS}$ and $(m-m_{\rm HB})_{\rm DECam}$ for the $gri$ magnitudes of the stars in the 19, 23 and 4 clusters in common in these filters, respectively.

\begin{figure*}[!htb]
   \centering
   \includegraphics[width=0.32\textwidth]{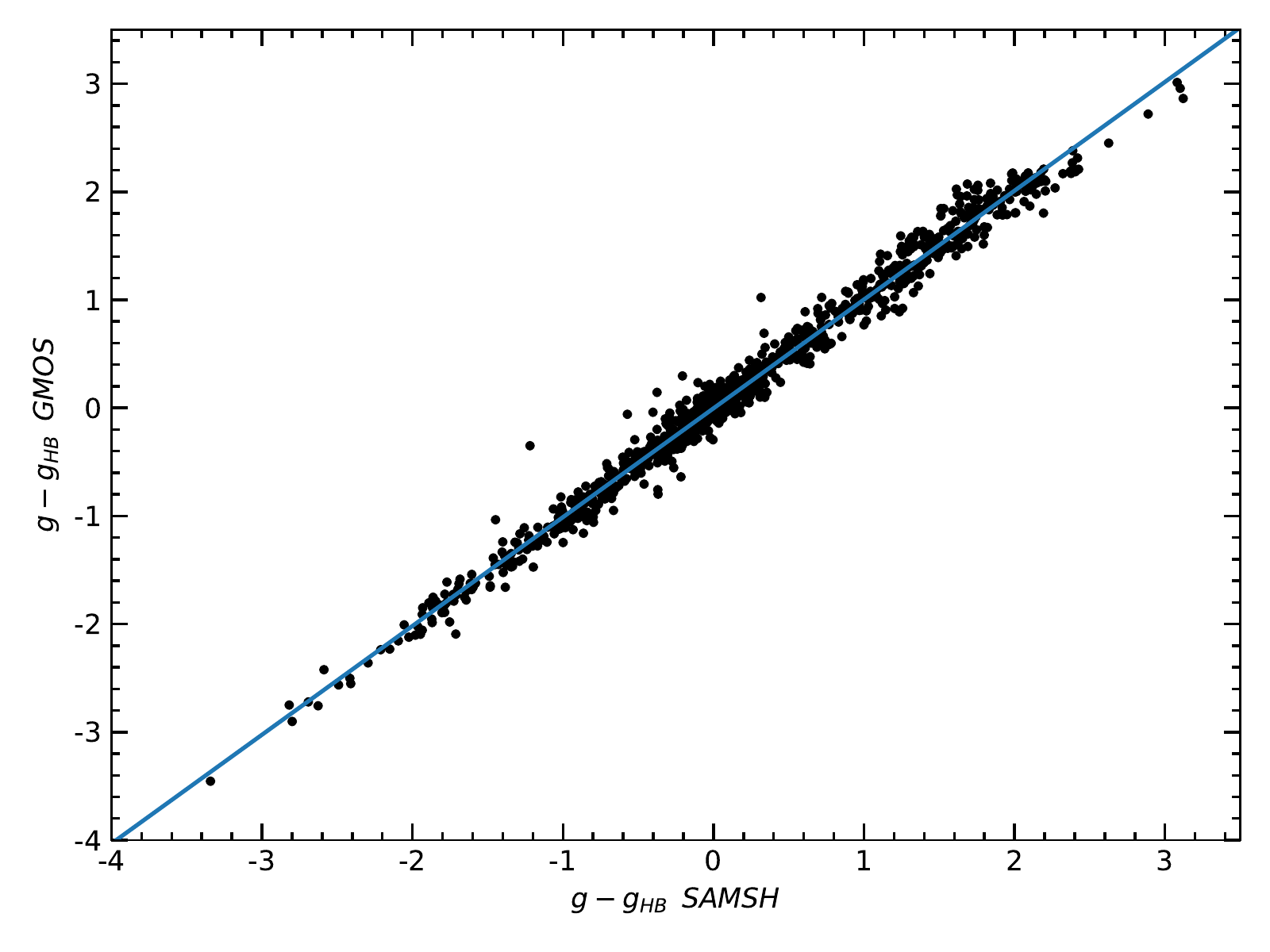}
   \includegraphics[width=0.32\textwidth]{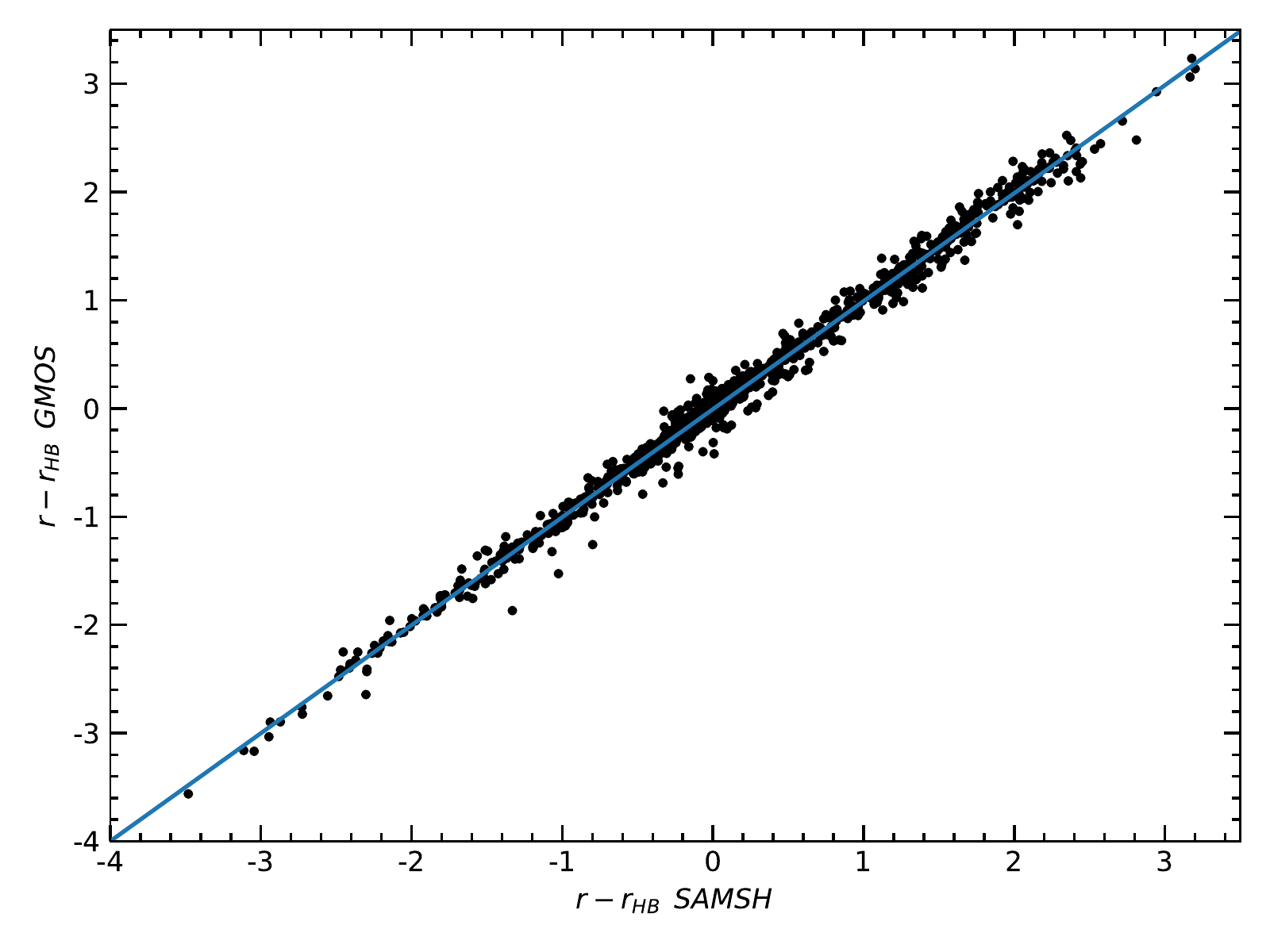}
   \includegraphics[width=0.32\textwidth]{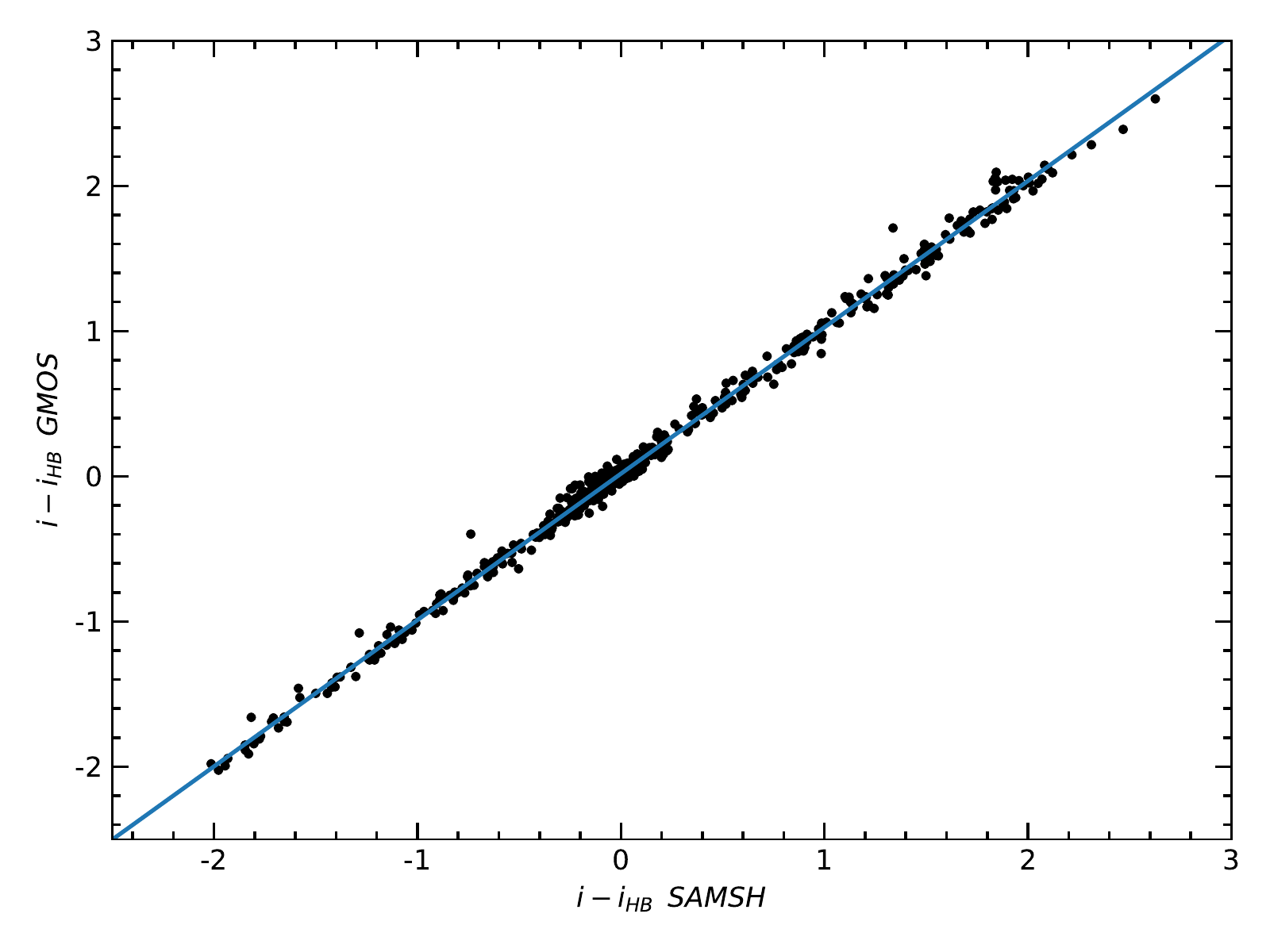}
      \caption{One-to-one relation between $(m-m_{\rm HB})_{\rm GMOS}$ and $(m-m_{\rm HB})_{\rm DECAM}$ for $gri$ filters with the respective linear fits.} 
         \label{fig:gmos-smash}
   \end{figure*}

Figure \ref{fig:gmos-smash} shows the fits with a very low dispersion in all cases. The results are displayed in Table \ref{tab:gmos-smash} from where it becomes clear that the slope is consistent with unity and the offset is consistent with zero within 1-2$\sigma$. In conclusion, the use of $(m-m_{\rm HB})_{\rm GMOS}$ and $(m-m_{\rm HB})_{\rm DECam}$ are interchangeable. Therefore, all the analysis presented in this paper using DECam $gri$ photometry is also valid for the GMOS $gri$ photometry. We note that our GMOS observations do not contain the $z$ filter, but we perform the analysis with the DECam $z$ filter for completeness.
Additionally, this one-to-one comparison is strictly valid for the relative magnitudes of RGB and RC stars; a proper conversion of magnitude systems between GMOS and DECam should take into account colour terms, for example.


\begin{table}[!htb]
\caption{$(m-m_{\rm HB})_{\rm GMOS}$ and $(m-m_{\rm HB})_{\rm DECam}$ linear fit results.}             
\label{tab:gmos-smash}      
\centering                          
\begin{tabular}{l c c c}        
\hline\hline                 
filter & slope & offset & r.m.s. \\    
\hline                        
g & 1.007$\pm$0.004  & -0.002$\pm$0.004 & 0.127 \\      
r & 0.998$\pm$0.003 & -0.007$\pm$0.003 &  0.102 \\
i & 1.008$\pm$0.003 & 0.018$\pm$0.03 & 0.058 \\
\hline                                   
\end{tabular}
\end{table}

\section{Reduced equivalent widths}

The concept of reduced equivalent widths ($W'$) was borrowed from stellar high-resolution spectroscopic analysis to be applicable to the CaT technique by AD91. At first glance, the aim is to normalise some extra effect on the EW of a line and leave its dependence on the chemical abundance of a given star. In the first case, the EW is divided by the wavelength to normalise Doppler-dependent phenomena \citep{gray08}; in the second case, the EW is compensated by the luminosity along the RGB, as given by Eq. \ref{eq:REWV}.
The motivation of AD91 was that they could directly compare the EW of stars with different luminosities within the same star cluster and detect, for example, an intrinsic metallicity spread in clusters \citep[e.g.][]{dacosta+09,dacosta+14}.

The $W'$ of a star cluster can be defined in two ways. The first way has three steps \citep[e.g.][]{saviane+12}: fit Eq. \ref{eq:REWV} to all member stars in a cluster to find $\beta$; apply $\beta$ in Eq. \ref{eq:REWV} to find $W'_i$ for each star; calculate the average $W' = \langle W_i'\rangle$. This is tantamount to the second way (e.g. C04, P09, P15), i.e., the representative $W'$ of a star cluster is simply the zero point of Eq. \ref{eq:REWV}. We note that the $W'$-[Fe/H] relation can be fitted using $W'$ or $\langle W_i'\rangle$ (see Sect. \ref{sec:catfeh}).  
As we are following the recipes of C04, P09, P15, we apply the second case here.

There are three quantities in Eq. \ref{eq:REWV} that have been exhaustively explored in the literature, namely, the measurement of the EW itself; the slope $\beta$ as a function of luminosity; the relation of the cluster $W'$ with metallicity. Some works have tried different filter choices, but no study has thoroughly investigated this matter, as we propose to do in this section.

For the EW definitions, some references are listed below with the relevant points, while a discussion on the difference of using the sum of all three CaT lines or just the sum of the two strongest lines is investigated in Sect. \ref{sec:2l3l}.
\begin{itemize}
    \item lines and continua bandpasses: many works have defined the windows for continuum and for the CaT line profile that were adapted essentially to resolution and metallicity, and many studies adopt their definitions \citep[e.g.][just to mention a few; see Table \ref{tab:betalit}]{armandroff+91,rutledge+97a,cenarro+01,battaglia+08,vasquez+15}. However, \cite{carrera+07} argued that $\sum EW$ does not depend on the definition of the bandpasses;
    \item line profile function: integrating the function that best fits the CaII line profile is usually preferred over the direct integration of the flux, as it reduces dependency on SNR and spurious weak lines depending on metallicity and spectral resolution. Nevertheless, for more metal-rich stars, the line is very saturated and the EW loses sensitivity in the core and the wings are more important; moreover, the continuum level is lower due to more free electrons in the atmosphere \citep[e.g.][]{erdelyi-mendes+91,vasquez+15}. Therefore, fitting a Gaussian does not account for the entire line of a more metal-rich star and a combination of Gaussian and Lorentzian function has to be used \citep[e.g.][]{saviane+12}, or some correction to the Gaussian EW \citep[e.g.][]{battaglia+08}, or some alternative function like Moffat \citep[e.g.][]{rutledge+97a}, as listed in Table \ref{tab:betalit}.
\end{itemize}

\begin{table*}[!htb]
\caption{Compilation of the main works on CaT metallicities with similar analysis done here, i.e., with $\beta$ available for comparison, sorted by publication year. Other details of each analysis are displayed in the other columns.}             
\label{tab:betalit}      
\centering                          
\footnotesize
\begin{tabular}{l l l l l l}        
\hline\hline                 
\noalign{\smallskip}  
  \multicolumn{1}{c}{ID} &
  \multicolumn{1}{c}{Function} &
  \multicolumn{1}{c}{Bandpass} &
  \multicolumn{1}{c}{$\sum$ EW} &
  \multicolumn{1}{l}{Luminosity} &
  \multicolumn{1}{l}{$\beta$} \\
\noalign{\smallskip}
\hline 
\noalign{\smallskip}
AD91            & G                     & AD91               & 2L              & $(V-V_{HB})$     & 0.62$\pm$0.01  \\
ADZ92$^a$       & G                     & AD91               & 2L              & $(V-V_{HB})$     & 0.66$\pm$0.07  \\
DAN92$^a$       & G                     & AD91               & 2L              & $(V-V_{HB})$     & 0.72$\pm$0.04  \\
S93             & G                     & AZ88               & 2L              & $(V-V_{HB})$     & 0.64$\pm$0.03  \\
G95             & N                     & AZ88               & 2L              & $(V-V_{HB})$     & 0.61$\pm$0.04  \\
DA95$^a$        & G                     & AD91               & 2L              & $(V-V_{HB})$     & 0.61$\pm$0.03  \\
SK96            & G                     & AZ88               & 2L              & $(V-V_{HB})$     & 0.62	 \\  
R97$^{a,b}$     & M                     & R97                & 3L*             & $(V-V_{HB})$     & 0.64$\pm$0.02  \\
T01$^b$         & G                     & ---                & 2L              & $(V-V_{HB})$     & 0.64$\pm$0.02  \\
C04$^b$         & G+L                   & AZ88               & 3L              & $(V-V_{HB})$     & 0.73$\pm$0.04  \\
P04$^b$        & G,M                   & R97                & 3L*               & M$_I$               &0.48$\pm$0.02 \\
Car07$^{a,b,c}$ & G+L                   & Cen01              & 3L              & M$_V$, M$_I$     & 0.677$\pm$0.004, 0.611$\pm$0.002\\  
B08$^b$         & G                     & B08                & 2L,3L,3L*       & $(V-V_{HB})$     & 0.62$\pm$0.03, 0.79$\pm$0.04, 0.59$\pm$0.04	\\
S12$^{b,c,d}$   & G,G+L                 & AD91               & 2L              & $(V-V_{HB})$     & 0.627           \\
M14$^{d,e}$     & \multicolumn{3}{c}{(Taken from S12 and R97)}    & $(K_s-K_{s,HB})$ & 0.385$\pm$0.013  \\
V15             & G+L                   & V15                & 2L              & $(K_s-K_{s,HB})$ &	0.384$\pm$0.019   \\
DC16            & G+L                   & AD91               & 2L              & $(V-V_{HB})$     & 0.66$\pm$0.016 \\
V18$^{f}$       & G+L                   & AD91               & 2L              & $(V-V_{HB})$     & 0.55	   \\
H20$^{f}$       & V                     & Cen01              & 2L              & $(F606W-F606W_{HB})$       & 0.581$\pm$0.004	  \\
\noalign{\smallskip}
\hline                                   
\end{tabular}
\tablefoot{
{\it ID, bandpasses.}
AD91: \cite{armandroff+91}; 
ADZ92: \cite{armandroff+92}; 
DAN92: \cite{dacosta+92}; 
S93: \cite{suntzeff+93}; 
G95: \cite{geisler+95}; 
DA95: \cite{dacosta+95}; 
SK96: \cite{suntzeff+96}; 
R97: \cite{rutledge+97b,rutledge+97a}; 
T01: \cite{tolstoy+01}; 
C04: \cite{cole+04}; 
P04: \cite{pont+04};
Car07: \cite{carrera+07}; 
B08: \cite{battaglia+08}; 
S12: \cite{saviane+12};  
M14: \cite{mauro+14}; 
V15: \cite{vasquez+15}; 
DC16: \cite{dacosta16}; 
V18: \cite{vasquez+18}; 
H20: \cite{husser+20};
AZ88: \cite{armandroff+88};
Cen01: \cite{cenarro+01}.\\
{\it Function}.
G: Gaussian;
G+L: Gaussian + Lorentzian;
M: Moffat;
V: Voigt;
N: Numerical.\\
{\it Metallicity scales}. 
(a) \cite{zinn+84}, 
(b) \cite{carretta+97}, 
(c) \cite{kraft+03}, 
(d) \cite{carretta+09}, 
(e) \cite{harris10}, 
(f) \cite{dias+16}.\\
* weighted.}
\end{table*}

With regard to the slope $\beta$, AD91 realised that $(V-V_{\rm HB})$ had a constant $\beta_{V} ^{AD91} = 0.631\pm0.018\  (\sigma=0.029)$\footnote{AD91 reported $\beta_{V} ^{AD91} = 0.619\pm0.010$, but if we take the average of the reported slopes from their Table 5, weighted by the inverse of the squared error, we find a slightly different slope and uncertainty.} with $\sum EW = EW_{8542} + EW_{8662}$ in the range -2.2 $<$ [Fe/H] $<$ -0.7 and $-3.0  < (V-V_{\rm HB}) < -0.4$, meaning that it would be possible to derive and compare $W'_i$ (and [Fe/H]) for any individual star $i$ within this parameter space. This statement has been questioned by many later works who found a variation of $\beta$ with metallicity and in some cases, even a non-linear relation from theory and data \citep[e.g.][]{pont+04,starkenburg+10,carrera+13,husser+20}.
In this work we are testing the effect of the choice of filters and the number of lines, taking as a test bed a sub-sample of SMC clusters stars of P09, P15 and the procedures of C04. Therefore, we assume $\beta_m$ fixed for all clusters as done in these reference works to avoid increasing the degrees of freedom in the analysis. In particular, we fit Eq. \ref{eq:REWV} for each cluster and calculate the average $\beta_m$ for the sample (see Table \ref{tab:slopes}), that is then fixed to find $W'_m$ for each cluster (see Table \ref{tab:W}). We leave the discussion on alternatives for Eq. \ref{eq:REWV} to the aforementioned works. 
The slope $\beta_m$ as a function of the filter $m$ is discussed in Sect. \ref{sec:beta}
The effects of the filter choice on the final $W'$ are analysed in Sect. \ref{sec:rew} and the $W'$-[Fe/H] relation is discussed in Sect.\ref{sec:catfeh} .

\subsection{Dependence of $\beta$ on the filter}
\label{sec:beta}

Equation \ref{eq:REWV} is fitted to all filters for all clusters individually and their slopes $\beta_m$ per filter are averaged to investigate the theoretical predictions from Sect. \ref{sec:isoc}. The results presented in Table \ref{tab:slopes} show similar errors as reported in the literature (reported in Table \ref{tab:betalit}), and there is also a non-negligible cluster-by-cluster dispersion expressed by the standard deviation in the table, that was already noted in Sect. \ref{sec:sample} and will not be further discussed here.

\begin{table}[!htb]
\caption{Weighted average of $\beta_m$ from fitting Eq. \ref{eq:REWV} to all clusters for each filter $m$, where $\sum EW = EW_{8498} + EW_{8542} + EW_{8662}$. Formal uncertainties and standard deviation are presented.}             
\label{tab:slopes}      
\centering                          
\begin{tabular}{l r c}        
\hline\hline                 
$m$ & $\lambda_{\rm eff}$ & $\langle\beta_{\rm 3L}\rangle$ \\    
 & (nm)  & $\pm$unc$\pm$std.dev. \\
\hline                        
$B$  & 445.958     & 0.84$\pm$0.12$\pm$0.33 \\
$g$  & 488.620     & 0.85$\pm$0.08$\pm$0.21 \\
$V$  & 550.002     & 0.71$\pm$0.05$\pm$0.14 \\
$G$  & 641.571     & 0.66$\pm$0.07$\pm$0.18 \\
$r$  & 642.833     & 0.67$\pm$0.07$\pm$0.19 \\
$i$  & 780.145     & 0.62$\pm$0.06$\pm$0.17 \\
$I$  & 801.020     & 0.58$\pm$0.07$\pm$0.18 \\
$z$  & 915.144     & 0.58$\pm$0.07$\pm$0.19 \\
$Y$  & 1\,023.964  & 0.57$\pm$0.06$\pm$0.15 \\           
$J$  & 1\,254.363  & 0.54$\pm$0.05$\pm$0.13 \\           
$Ks$ & 2\,134.285  & 0.48$\pm$0.06$\pm$0.15 \\           
\hline                                   
\end{tabular}
\end{table}

\begin{figure}[!htb]
\centering
\includegraphics[width=\columnwidth]{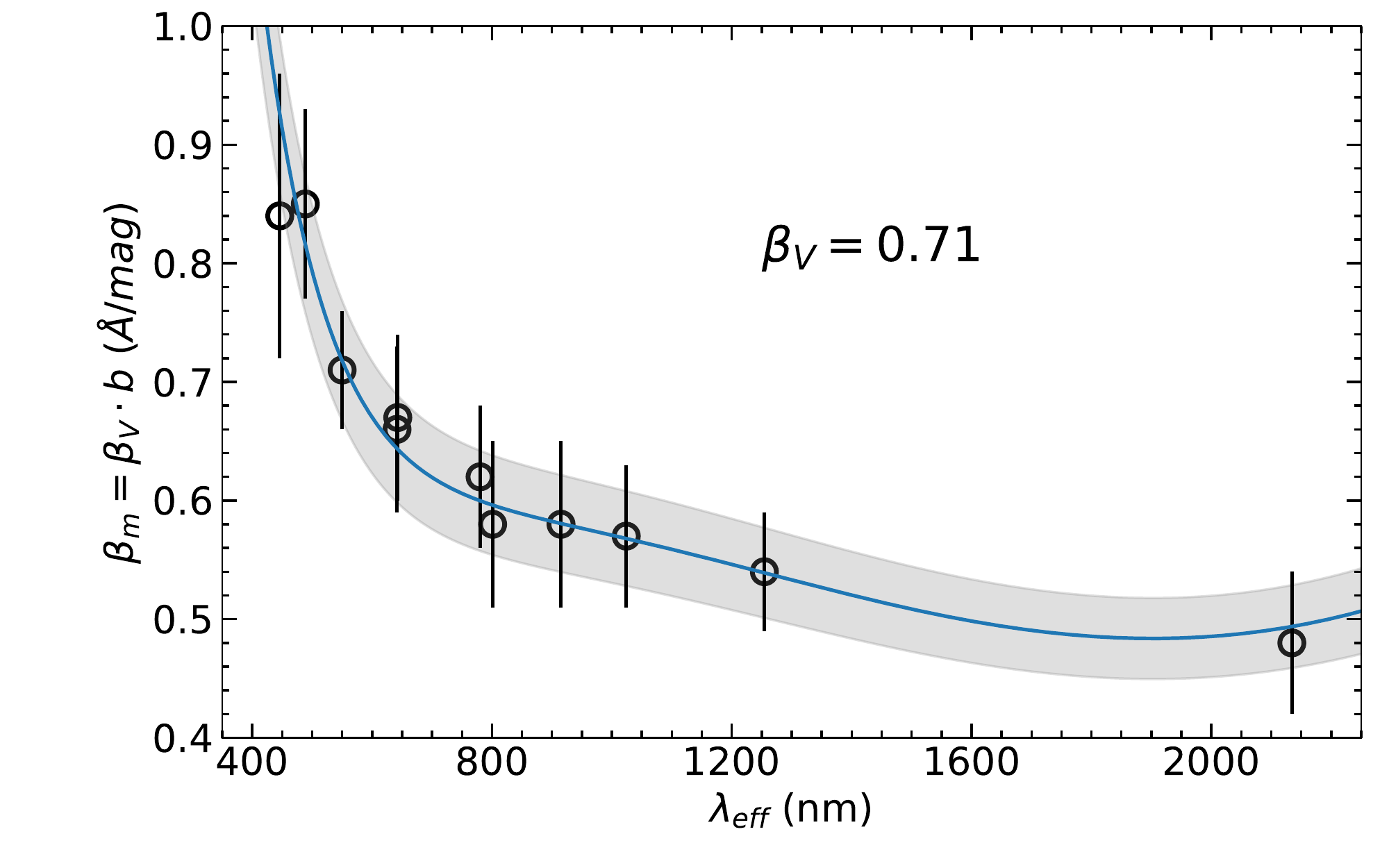}
   \caption{Average $\langle \beta_m \rangle$ for the seven clusters for each filter $m$ from Tab. \ref{tab:slopes}. The line is based on Eq.\ref{eq:b-poly} and the shaded area displays the $1\sigma$ around the function, which is dominated by the error in $\beta_V$.}
      \label{fig:isoc-clusters}
\end{figure}

The results from Table \ref{tab:slopes} are compared to the theoretical predictions from Sect. \ref{sec:isoc} in Fig. \ref{fig:isoc-clusters}. The figure is based on Fig. \ref{fig:isoctypical} now showing explicitly $\beta_m$ anchoring the curves in the fitted $\beta_V = 0.71$. There is a very good agreement of the fitted $\beta_m$ with the predictions from the isochrones within less than $1\sigma$. Therefore, Eq. \ref{eq:b-poly} can be used to predict $\beta_m$ for a given $\beta_V$. Applying this conclusion to another case, \cite{saviane+12} and \cite{vasquez+18} use $m=V$ and have applied the same analysis criteria to find $\beta_V = 0.627$ and $0.55$, respectively. On the other hand, \cite{mauro+14} follows the same criteria but using $m = K_s$ instead, as it was also done by \cite{vasquez+15}, who found $\beta_{K_s} = 0.384\pm0.019$ and $0.385\pm0.013$, respectively (see Tab. \ref{tab:betalit}). Using Eq.\ref{eq:b-poly} with $\beta_V$ aforementioned, the predictions for $\beta_{K_s}$ are 0.44 and 0.38, respectively, in agreement with the fitted slopes. Another point that calls the attention in Fig. \ref{fig:isoc-clusters} is that filters $r$ and $G$ have similar behaviour as they have similar $\lambda_{eff}$ (see Tab. \ref{tab:slopes}), even though the width of their transmission curves are very different (see Fig. \ref{fig:filters}), showing that the relevant characteristic of a filter is the $\lambda_{eff}$. 

We note that filters that are bluer than the reference $V$ present larger discrepancies between prediction and data in Fig. \ref{fig:isoc-clusters}. Therefore, $V$ filter or redder alternatives should be preferred. We conclude that it is possible to take an existent CaT calibration using $V$ filter and assume the theoretical prediction of $\beta_m$ with an average error of $0.05\ \AA/mag$ for a redder filter and simply swap the filters in the CaT analysis that will continue valid.

\begin{figure}[!htb]
\centering
\includegraphics[width=\columnwidth]{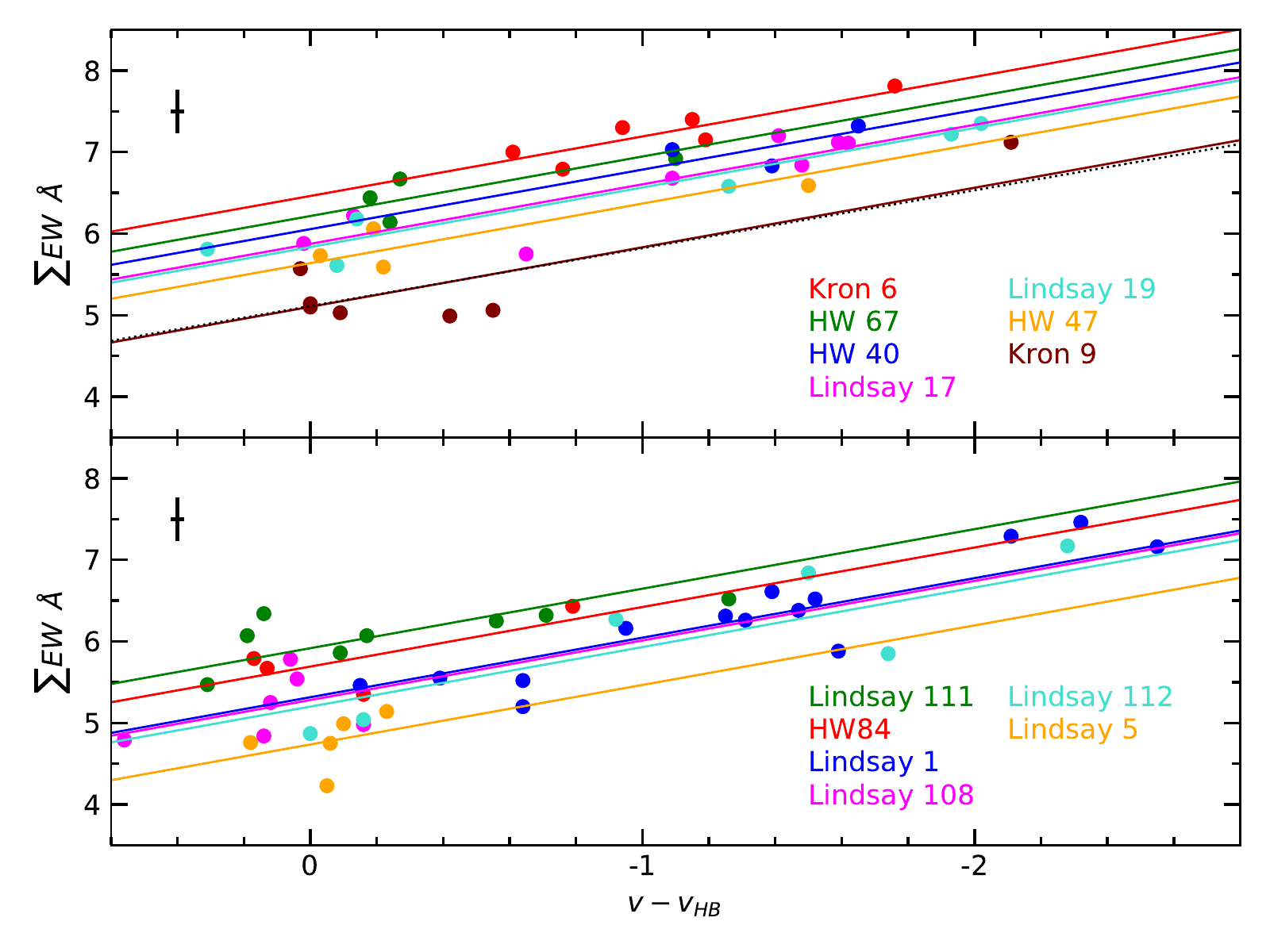}
   \caption{Results of $\sum$EW vs. mag difference fitting with 3 lines and $V$ filter for each cluster keeping fixed the slope $\beta_V=0.73\pm0.04\ \AA/mag$ from the extended sample of 13 clusters. The top panel shows the selected sample of seven clusters and the dashed line shows the fitted $\beta_V = 0.71\pm0.05\ \AA/mag$ (see Table \ref{tab:slopes}) for this sample to make it clear that the difference is negligible. Typical error bars are indicated on the top left corner. Equivalent figures for the other filters are presented in the Appendix.
   }
      \label{fig:WV}
\end{figure}

As discussed in Sect. \ref{sec:sample}, the sample analysed here spans the same range in $\beta_V$ as the sample of C04 with the goal to reproduce their $\beta_V = 0.73\pm0.04\ \AA/mag$. We found $\beta_V = 0.71\pm0.05\ \AA/mag$ which is in very good agreement, as expected. We note that if instead of applying all selection criteria described in Sect. \ref{sec:sample} and consider only clusters with $\beta_V$ within the range found by C04, the sample increases to 13 clusters and the average is $\beta_V=0.73\pm0.04\ \AA/mag$, i.e., the sample size does not change $\beta$, but we keep the original sample of seven clusters to be able to derive $\beta_m$ for all other filters. For a graphical representation, we show the fits for the sample of 13 clusters and for the adopted sample of seven clusters in Fig. \ref{fig:WV} from where it is clear that the points of a given cluster are represented well by the fit with a fixed $\beta_V$. This was already shown in the original works by P09, P15, but we reproduce the analysis here to setup the reference of the analysis using the $V$ filter.

We have made a compilation of CaT works in the literature that have similar analysis as that presented in this work, i.e., that publish a $\beta$ that can be compared (see Tab. \ref{tab:betalit}). This is not a complete list, but it contains a vast record of $\beta$ for a variation of criteria, and covering publication years from 1991 to 2020. It is noticeable that the average $\beta_V = 0.642\pm0.057\ \AA/mag$ is very stable among the works, even though there are works using $\beta_V < 0.6$ and $\beta_V > 0.7$. Our derived $\beta_V = 0.71\pm0.03\ \AA/mag$ agrees with the average, and it is not the only $\beta > 0.7$; \cite{dacosta+92} and \cite{battaglia+08} derived $\beta_V = 0.72\pm0.04\ \AA/mag$ and $0.79\pm0.04\ \AA/mag$, respectively, not to mention C04.

\subsection{Dependence of $W'$ on the filter}
\label{sec:rew}

A prime goal of this work is to answer the question whether the filter choice on Eq. \ref{eq:REWV} makes any difference in the final metallicity, i.e., on $W'$, and if so by how much. Fig. \ref{fig:isoctypical} shows that for all filters 
$\langle W'_m\rangle \approx W'_V\pm0.15\ \AA/mag$, i.e., $W'$ is constant with filter within typical uncertainties.

\begin{figure}[!htb]
   \centering
   \includegraphics[width=\columnwidth]{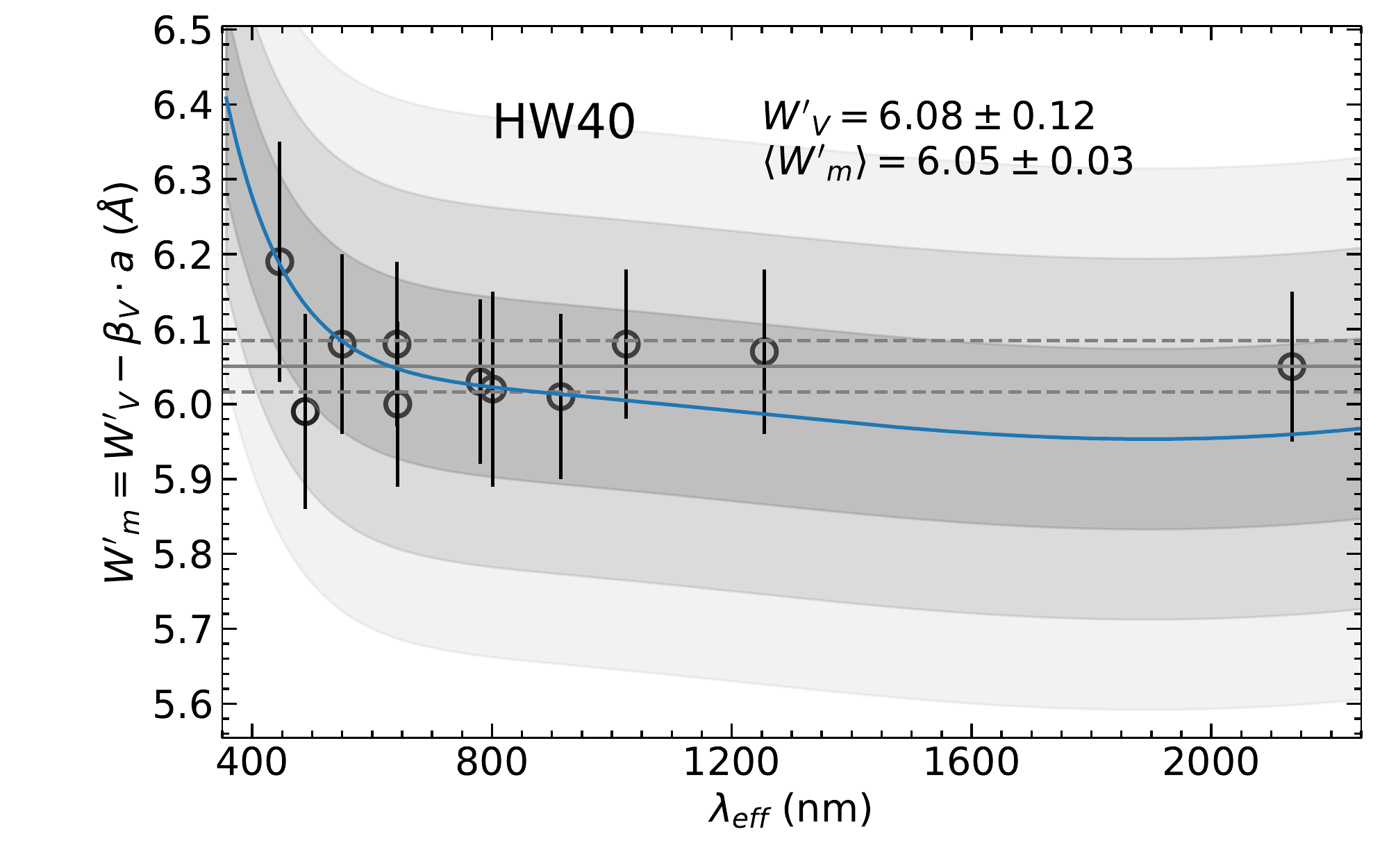}
      \caption{Reduced equivalent width $W'_m$ for the cluster HW\,40 for each filter $m$. The grey solid and dashed lines are the average $\langle W'_m \rangle$ and respective error (see Tab. \ref{tab:W}).
      The solid blue line is based on Eq. \ref{eq:a-poly} and $\beta_V = 0.71$ from Tab. \ref{tab:slopes}, and the respective shaded areas represent $\pm1,2,3\sigma$ around the function, which is dominated by the error in $W'_V$. Equivalent figures for the other clusters are presented in the Appendix.}
         \label{fig:HW40Wall}
   \end{figure}

As described in the previous subsection, the final fits represented in Fig. \ref{fig:WV} is made with a fixed averaged $\langle\beta_m\rangle = 0.71\pm0.05\ \AA/mag$, using Eq. \ref{eq:REWV}. The zero point $W'_m$ and its associated error comes out from this linear fit.
The resulting $W'_m$ for all clusters and all filters are presented in Table \ref{tab:W}. It is clear that $W'$ remains unchanged within uncertainties regardless the filter choice in Eq. \ref{eq:REWV}. 
A visual verification of these results is shown in Fig. \ref{fig:HW40Wall} for the cluster HW\,40, as an example. In general, the measured $W'_m$ agrees within $1\sigma$ with the predicted values for filters $V$ or redder than that. 
The conclusion is the same as for the $\beta_m$ analysis, i.e., redder filters are preferred. Also, the resulting $W'_m$ for a particular filter $m$ is consistent with $W'_V$ within the average error of $0.15\ \AA$, for a given CaT calibration.

\begin{table*}
\caption{$W'_m$ for all clusters in every filter $m$ using 3 lines, with its respective uncertainty and r.m.s. error. The average $\langle W'_m \rangle$ of all filters for each cluster is also presented.}             
\label{tab:W}      
\centering                          
\footnotesize
\begin{tabular}{lrrrr}
\hline
  \multicolumn{1}{c}{Cluster} &
  \multicolumn{1}{c}{$W'_B$} &
  \multicolumn{1}{c}{$W'_g$} &
  \multicolumn{1}{c}{$W'_V$} &
  \multicolumn{1}{c}{$W'_G$} \\
    \multicolumn{1}{c}{} &
  \multicolumn{1}{c}{$W'_r$} &
  \multicolumn{1}{c}{$W'_i$} &
  \multicolumn{1}{c}{$W'_I$} &
  \multicolumn{1}{c}{$W'_z$} \\
  \multicolumn{1}{c}{} &
  \multicolumn{1}{c}{$W'_Y$} &
  \multicolumn{1}{c}{$W'_J$} &
  \multicolumn{1}{c}{$W'_{Ks}$} &
  \multicolumn{1}{c}{$\boldsymbol{\langle W'_m \rangle}$} \\
\hline 
\noalign{\smallskip}
Kron\,6      & 6.40$\pm$0.17$\pm$0.27 & 6.39$\pm$0.05$\pm$0.13 & 6.48$\pm$0.07$\pm$0.16 & 6.48$\pm$0.07$\pm$0.16 \\
           & 6.39$\pm$0.06$\pm$0.14 & 6.43$\pm$0.06$\pm$0.16 & 6.48$\pm$0.11$\pm$0.27 & 6.45$\pm$0.07$\pm$0.17  \\
           & 6.41$\pm$0.09$\pm$0.22  & 6.45$\pm$0.09$\pm$0.22 & 6.44$\pm$0.09$\pm$0.21 &\bf{6.43$\pm$0.02$\pm$0.03}\\
\noalign{\smallskip}
HW\,67       & 5.95$\pm$0.09$\pm$0.18 & 6.17$\pm$0.12$\pm$0.23 & 6.22$\pm$0.11$\pm$0.22 & 6.19$\pm$0.11$\pm$0.22  \\
           & 6.15$\pm$0.11$\pm$0.22 & 6.18$\pm$0.11$\pm$0.22 & 6.12$\pm$0.10$\pm$0.19 & 6.19$\pm$0.11$\pm$0.22  \\
           & 6.18$\pm$0.11$\pm$0.21  & 6.22$\pm$0.11$\pm$0.21  & 6.21$\pm$0.11$\pm$0.22 &\bf{6.15$\pm$0.03$\pm$0.07}\\
\noalign{\smallskip}
HW\,47       & 5.59$\pm$0.11$\pm$0.22 & 5.65$\pm$0.11$\pm$0.23 & 5.65$\pm$0.11$\pm$0.22 & 5.71$\pm$0.09$\pm$0.18  \\
           & 5.68$\pm$0.10$\pm$0.20 & 5.72$\pm$0.10$\pm$0.20 & 5.61$\pm$0.19$\pm$0.38 & 5.69$\pm$0.10$\pm$0.19  \\
           & 5.61$\pm$0.14$\pm$0.27  & 5.65$\pm$0.11$\pm$0.21 & 5.69$\pm$0.10$\pm$0.19&\bf{5.67$\pm$0.03$\pm$0.04}\\
\noalign{\smallskip} 
HW\,40       & 6.19$\pm$0.16$\pm$0.27 & 5.99$\pm$0.13$\pm$0.22 & 6.08$\pm$0.12$\pm$0.21 & 6.08$\pm$0.11$\pm$0.19  \\
           & 6.00$\pm$0.11$\pm$0.20 & 6.03$\pm$0.11$\pm$0.19 & 6.02$\pm$0.13$\pm$0.23 & 6.01$\pm$0.11$\pm$0.18  \\
           & 6.08$\pm$0.10$\pm$0.18  & 6.07$\pm$0.11$\pm$0.18 & 6.05$\pm$0.10$\pm$0.17 &\bf{6.05$\pm$0.03$\pm$0.05}\\
\noalign{\smallskip} 
Lindsay\,17  & 6.00$\pm$0.10$\pm$0.29 & 5.89$\pm$0.09$\pm$0.27 & 5.89$\pm$0.10$\pm$0.28 & 5.89$\pm$0.10$\pm$0.28  \\
           & 5.91$\pm$0.10$\pm$0.28 & 5.91$\pm$0.10$\pm$0.29 & 5.94$\pm$0.10$\pm$0.29 & 5.89$\pm$0.10$\pm$0.29  \\
           & 5.88$\pm$0.10$\pm$0.29  & 5.92$\pm$0.11$\pm$0.30 & 5.88$\pm$0.11$\pm$0.31 &\bf{5.91$\pm$0.03$\pm$0.03}\\
\noalign{\smallskip} 
Kron\,9      & 4.81$\pm$0.16$\pm$0.42  & 4.84$\pm$0.18$\pm$0.46 & 5.11$\pm$0.15$\pm$0.38 & 5.03$\pm$0.18$\pm$0.49  \\
           & 5.01$\pm$0.18$\pm$0.48  & 5.08$\pm$0.17$\pm$0.46 & 5.05$\pm$0.17$\pm$0.45 & 5.08$\pm$0.19$\pm$0.50  \\
           & 5.12$\pm$0.17$\pm$0.45  & 5.16$\pm$0.17$\pm$0.45 & 5.17$\pm$0.19$\pm$0.51 & \bf{5.04$\pm$0.05$\pm$0.11}\\
\noalign{\smallskip} 
Lindsay\,19  & 6.05$\pm$0.18$\pm$0.43  & 5.98$\pm$0.17$\pm$0.42 & 5.85$\pm$0.08$\pm$0.20 & 6.00$\pm$0.16$\pm$0.40   \\
           & 5.97$\pm$0.17$\pm$0.41  & 5.96$\pm$0.16$\pm$0.40 & 5.97$\pm$0.17$\pm$0.42 & 5.96$\pm$0.16$\pm$0.40  \\
           & 5.95$\pm$0.16$\pm$0.40  & 5.97$\pm$0.16$\pm$0.40 & 5.94$\pm$0.16$\pm$0.39 &\bf{5.94$\pm$0.04$\pm$0.05}\\
\noalign{\smallskip} 
\hline\end{tabular}
\end{table*}

\subsection{Dependence of $\beta$ and $W'$ on $\sum EW$ with 2 or 3 CaT lines}
\label{sec:2l3l}

On the top of all assumptions and definitions about Eq. \ref{eq:REWV}, we have been using $\sum EW = EW_{8498} + EW_{8542} + EW_{8662}$ to follow C04, P09, P15. However, many works use $\sum EW = EW_{8542} + EW_{8662}$ with the argument that the bluest and weakest line has lower SNR than the other stronger CaII lines and could introduce more noise than information to the analysis depending on the quality of the spectra. Naturally, $W'$ will be different for the two cases, therefore we analyse this difference here.

\begin{figure}[!htb]
\centering
\includegraphics[width=\columnwidth]{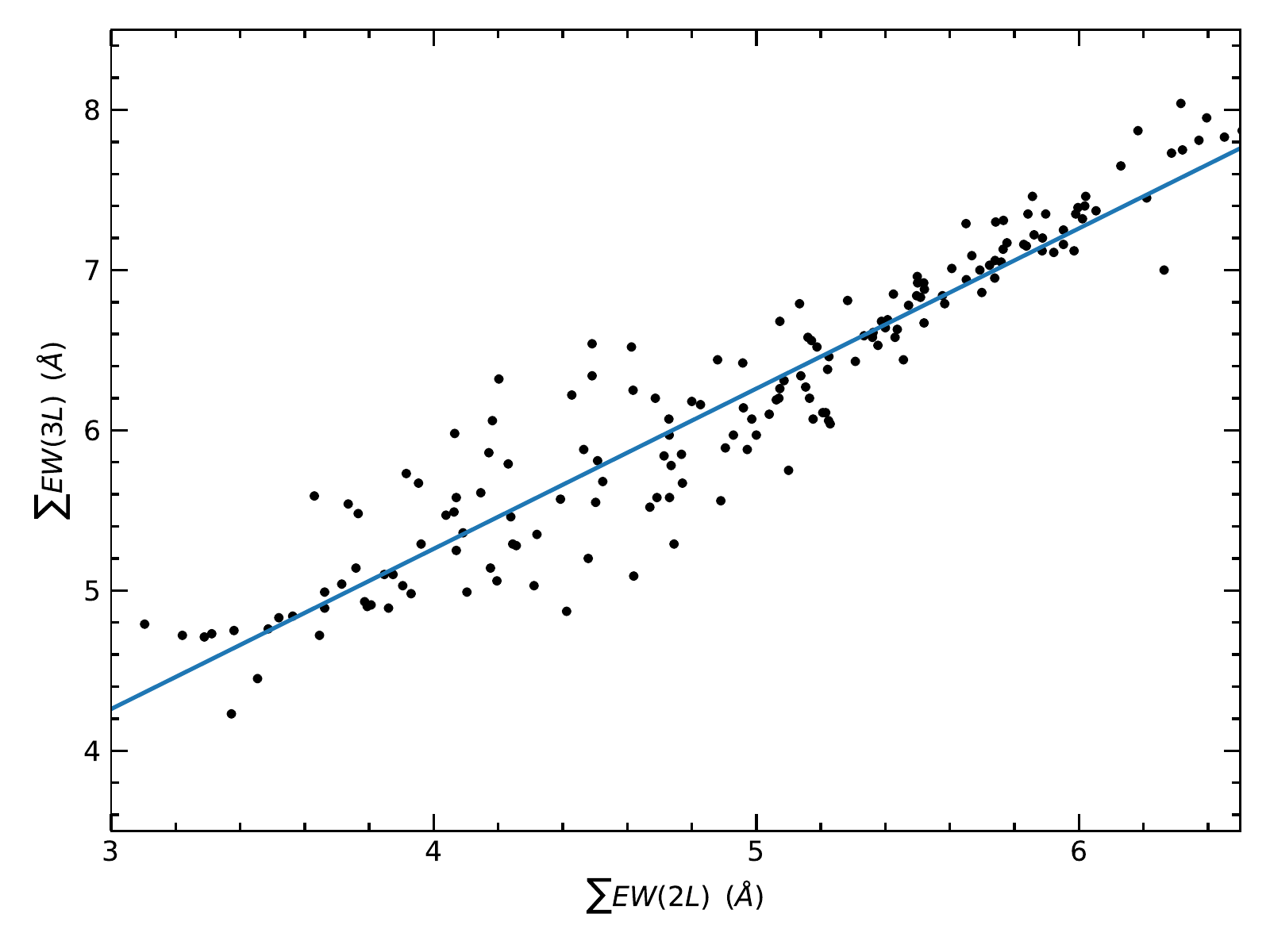}
   \caption{Relationship between the $\sum EW_{2L}$ and $\sum EW_{3L}$ for all stars from P09, P15 for reference. 
   }
      \label{fig:SEW2l3l}
\end{figure}

A good advantage of the CaT technique is that the sum of the EW of the three CaII lines or of the two strongest lines are proportional and therefore scalable. We found the relation

\begin{equation}
    \sum EW_{3L} = 1.26(\pm0.13) + 1.00(\pm0.03)\cdot \sum EW_{2L},
    \label{eq:2l3l}
\end{equation}

\noindent with an r.m.s. dispersion of $0.29\ \AA$ (see Fig. \ref{fig:SEW2l3l}). 

Similarly, there are a few works that use a weighted sum of the $EW$ of the three lines (see Tab. \ref{tab:betalit}) as a way of minimising the error in $\sum EW$ \citep[see discussion by][for example]{rutledge+97a}. Repeating the exercise performed above for two lines, and adopting $\sum EW_{3L*} = 0.5\cdot EW_{8498} + 1.0\cdot EW_{8542} + 0.6\cdot EW_{8662}$ from \cite{rutledge+97a}, we find the relation

\begin{equation}
    \sum EW_{3L} = 1.12(\pm0.19) + 1.00(\pm0.09)\cdot \sum EW_{3L*}.
    \label{eq:3l3lw}
\end{equation}

A direct consequence of Eqs. \ref{eq:2l3l} and \ref{eq:3l3lw} is that they can replace $\sum EW$ in Eq. \ref{eq:REWV}. The fact of the slope being unity simplifies the relations.
Replacing Eq. \ref{eq:REWV} into Eq. \ref{eq:2l3l} and Eq. \ref{eq:3l3lw}, we find that $\beta_m^{2L} = \beta_m^{3L}/A$ and $\beta_m^{3L*} = \beta_m^{3L}/A*$, where $A=A*=1.00$ is the slope from Eqs. \ref{eq:2l3l} and \ref{eq:3l3lw}.  In fact, we found $\beta_V^{2L} = 0.75\pm0.09\ \AA/mag$ and $\beta_V^{3L*} = 0.74\pm0.07\ \AA/mag$, which are both consistent with $\beta_m^{3L} = 0.71\pm0.05\ \AA/mag$.

\section{Ca II triplet metallicity calibration}
\label{sec:catfeh}

Having confirmed the predictions from Sect. \ref{sec:isoc} for $\beta_m$ and $W'_m$ using data from P09, P15, we now proceed to 
obtain final metallicities for the sample clusters and estimate their systematic uncertainties. We showed in Eq. \ref{eq:a-poly}, Figs. \ref{fig:isoc-clusters}, \ref{fig:HW40Wall} that the predicted variation in $W'_m$ with respect to $W'_V$ is of the order of $\Delta W' \lesssim \pm0.14\ \AA/mag$. We use the relation derived by C04,

\begin{equation}
    {\rm [Fe/H]} = -2.966 (\pm0.032) + 0.362 (\pm0.014)\cdot W',
    \label{eq:cat-feh}
\end{equation}

\noindent to convert this variation into $\Delta{\rm [Fe/H]} \lesssim \pm0.05\ dex$, which is within the typical error bars for CaT metallicities of about 0.1-0.2\ dex (e.g. P09, P15). This total uncertainty depends on the uncertainties from Eqs. \ref{eq:cat-feh} and \ref{eq:REWV}, and the largest contribution comes from Eq. \ref{eq:cat-feh}. In other words, there is an intrinsic dispersion on the $W'-{\rm [Fe/H]}$ relation that limits the uncertainty in the final CaT metallicity to about 0.1-0.2 dex for a given star.

\begin{figure}[!htb]
\centering
\includegraphics[width=\columnwidth]{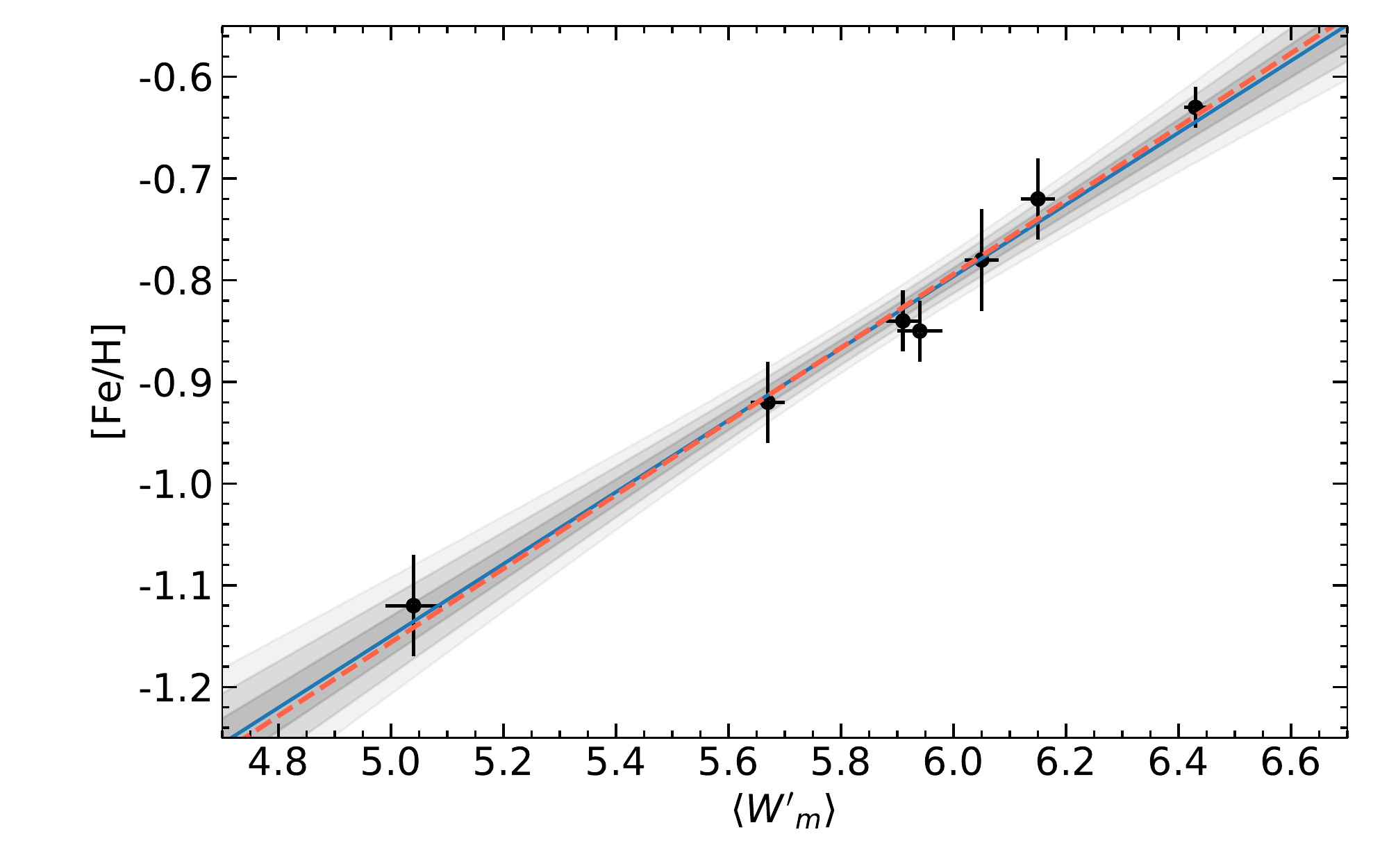}
   \caption{$\langle W'_m \rangle$-[Fe/H] relation using values from Tables \ref{tab:sample} and \ref{tab:W}. The solid blue line and respective shaded areas represent the linear fit and the $\pm1,2,3\sigma$ around the function. The orange dashed line represents the relation from C04 for reference.}
      \label{fig:cat-feh}
\end{figure}

Figure \ref{fig:cat-feh} displays the reference [Fe/H] from Tab.\ref{tab:sample} as a function of the average $\langle W'_m \rangle$ from Table \ref{tab:W} for each cluster. We performed a straight line fitting resulting in

\begin{equation}
    {\rm [Fe/H]} = -2.917 (\pm0.116) + 0.353 (\pm0.020)\cdot \langle W'_m \rangle,
    \label{eq:cat-fehnew}
\end{equation}

\noindent which is very good agreement with Eq. \ref{eq:cat-feh}, as also displayed in Fig. \ref{fig:cat-feh}.

Once again, if the same line profile function, number of CaT lines is used, the relation between $W'$ and [Fe/H] is the same regardless the filter used, producing [Fe/H] with a systematic error due to filter choice smaller than $\pm0.05$ dex. In other words, it is possible to use Eq.\ref{eq:cat-feh} from C04 with $W'$ derived with any filter, without adding a significant systematic error in the final metallicity.

\section{Summary and conclusions}

In this paper we investigate the effect of the choice of the luminosity indicator $(m-m_{HB})$ on the [Fe/H] derived from CaT for different filters $BVI,griz,G,YJK_s$. The EW of CaT lines are mostly sensitive to surface gravity and metallicity, therefore, a luminosity indicator is employed to remove the gravity effect leaving only the metallicity dependence. The effect generated by the choice of different filters can be predicted using isochrones.

The main conclusions from this work are listed below:
\begin{enumerate}
    \item The predictions of $\beta_m$ as a function of $\lambda_{eff}$ are confirmed by the data. The theoretical relation can be used to predict $\beta_m$ based on a given $\beta_V$ with an intrinsic accuracy of $\sim 0.05\ \AA/mag$;
    
    \item The predictions of $W'_m$ as a function of $\lambda_{eff}$ are confirmed by the data. The theoretical relation can be used with an intrinsic accuracy of $\sim 0.15\ \AA$ in $W'$, which means $\sim 0.05$ dex in [Fe/H];
    
    \item The average $\langle W'_m \rangle$ from all filters $m$ for each cluster follows the C04 relation with [Fe/H] with systematic errors smaller than 0.05 dex. Therefore, C04 scale can be used directly with $W'_m$ derived with any filter;
    
    \item The filters $V$ or redder tend to generate more stable results and leads to smaller systematic effects in the final [Fe/H].
\end{enumerate}    
    
The analysis in this work was performed for a limited range in metallicity and age, and for a sample of clusters from the SMC, following the recipes by C04. We showed that our predictions for $\beta$ are also consistent among the works by \cite{saviane+12,mauro+14,vasquez+15,vasquez+18} who studied old globular clusters and bulge stars in the Milky Way covering a wide range of metallicities. As a consequence, we speculate that the predictions based on the isochrones are applicable to any CaT analysis, and could be adapted even to other works that use more complex relations than simply linear functions.

Last, but not least, this work has direct implications to any CaT metallicity analysis for stars with panchromatic data available, with especial attention to the star clusters on Magellanic Clouds that have photometry from the surveys MCPS ($BVI$), VISCACHA ($BVI$), VISCACHA-GMOS ($gri$), SMASH ($griz$), Gaia ($G$), VMC ($YJK_s$), and others. These studies already started their spectroscopic follow-up or are about to start, using GMOS, 4MOST, APOGEE etc., in some cases covering the CaT.

\begin{acknowledgements}
This research was partially supported by the Argentinian institutions CONICET, SECYT (Universidad Nacional de Córdoba) and Agencia Nacional de Promoción Científica y Tecnológica (ANPCyT). \\
Based on observations collected at the European Southern Observatory under ESO programmes 076.B-0533, 082.B-0505 and 384.B-0687, 
and 179.B-2003. 
This research has made use of the services of the ESO Science Archive Facility.\\
This work has made use of data from the European Space Agency (ESA) mission
{\it Gaia} (\url{https://www.cosmos.esa.int/gaia}), processed by the {\it Gaia}
Data Processing and Analysis Consortium (DPAC,
\url{https://www.cosmos.esa.int/web/gaia/dpac/consortium}). Funding for the DPAC
has been provided by national institutions, in particular the institutions
participating in the {\it Gaia} Multilateral Agreement.\\
Based on observations obtained at the international Gemini Observatory, a program of NSF’s NOIRLab, which is managed by the Association of Universities for Research in Astronomy (AURA) under a cooperative agreement with the National Science Foundation. on behalf of the Gemini Observatory partnership: the National Science Foundation (United States), National Research Council (Canada), Agencia Nacional de Investigaci\'{o}n y Desarrollo (Chile), Ministerio de Ciencia, Tecnolog\'{i}a e Innovaci\'{o}n (Argentina), Minist\'{e}rio da Ci\^{e}ncia, Tecnologia, Inova\c{c}\~{o}es e Comunica\c{c}\~{o}es (Brazil), and Korea Astronomy and Space Science Institute (Republic of Korea). Program ID: GS-2017B-Q-19, GS-2018B-Q-208, GS-2018B-Q-302, and GS-2019B-Q-303.\\
This research uses services or data provided by the Astro Data Lab at NSF's National Optical-Infrared Astronomy Research Laboratory. NSF's OIR Lab is operated by the Association of Universities for Research in Astronomy (AURA), Inc. under a cooperative agreement with the National Science Foundation.\\
This paper includes data gathered with the 1.0 meter Swope Telescope located at Las Campanas Observatory, Chile.\\
We thank the referee for his/her analysis of our work, which has contributed to the improvement of the present study.

\end{acknowledgements}

   \bibliographystyle{aa} 
   \bibliography{bibliography} 

\appendix

\section{$\sum EW$ - magnitude dependence}

\begin{figure}[!htb]
   \centering
   \includegraphics[width=0.9\columnwidth]{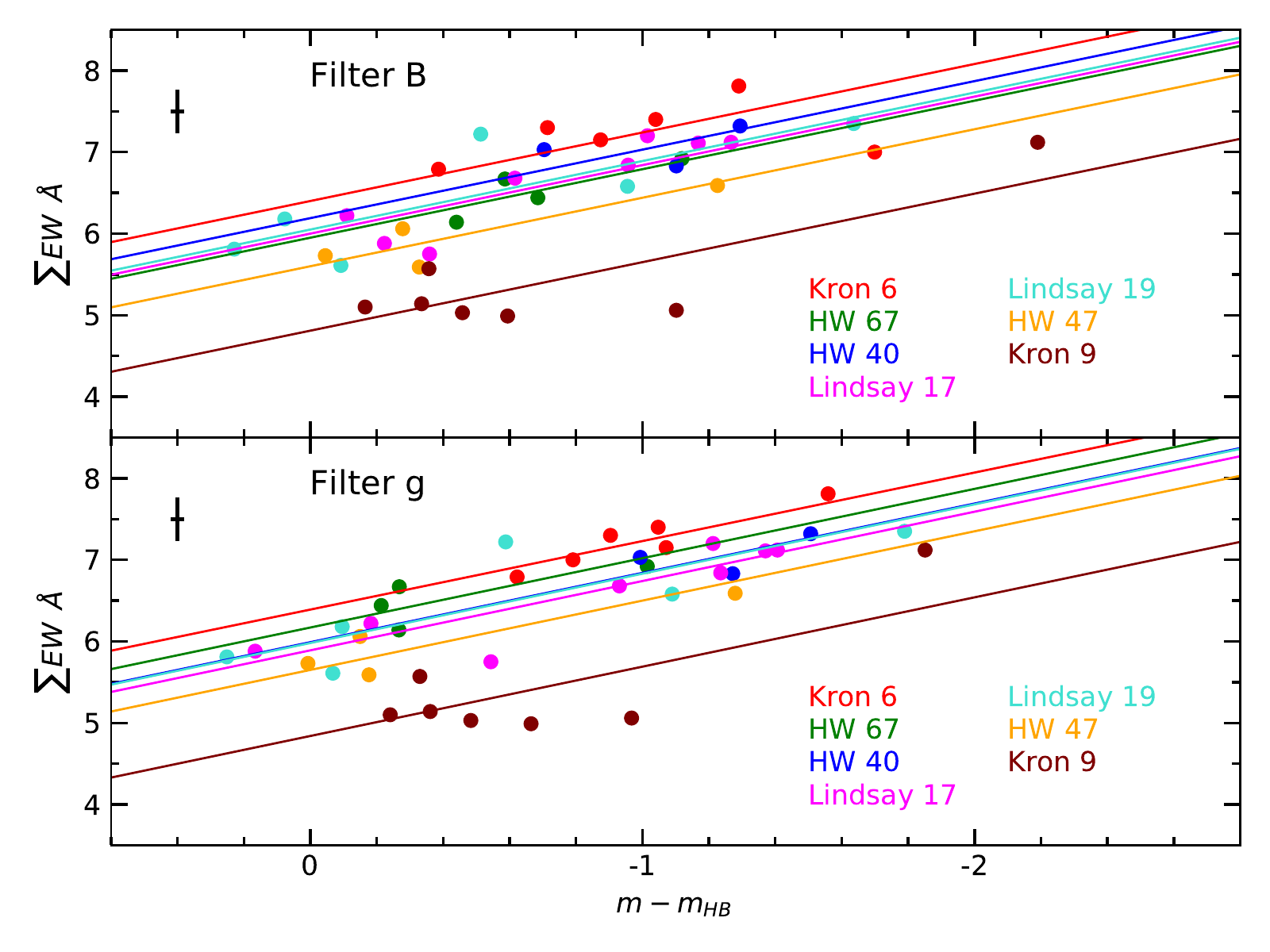}
      \caption{Same as Fig. \ref{fig:WV} for the $B$ and $g$ filters.}
         \label{fig:Wg}
   \end{figure}

\begin{figure}[!htb]
   \centering
   \includegraphics[width=0.9\columnwidth]{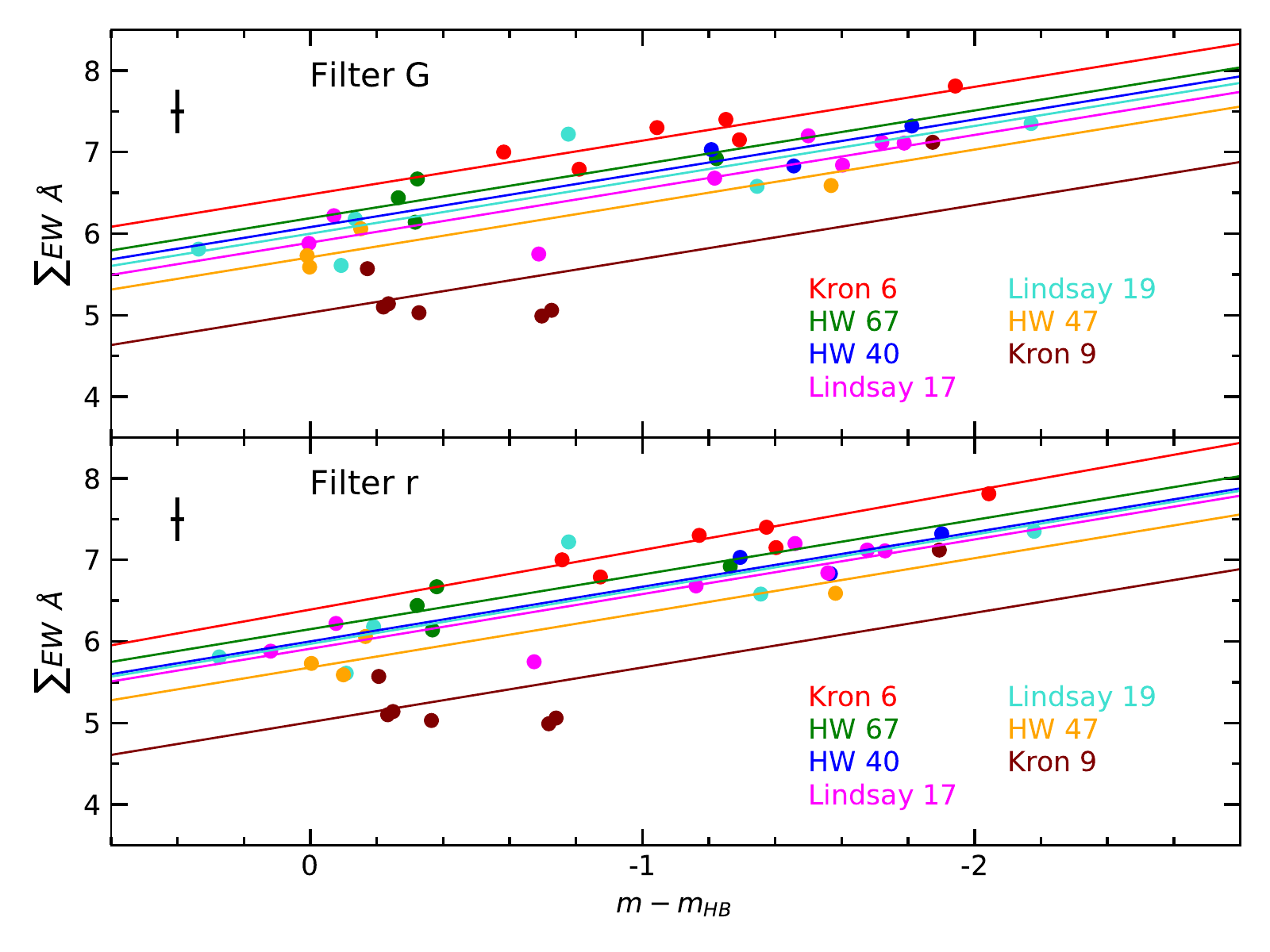}
      \caption{Same as Fig. \ref{fig:WV} for the $G$ and $r$ filters.}
         \label{fig:Wr}
   \end{figure}

\begin{figure}[!htb]
   \centering
   \includegraphics[width=0.9\columnwidth]{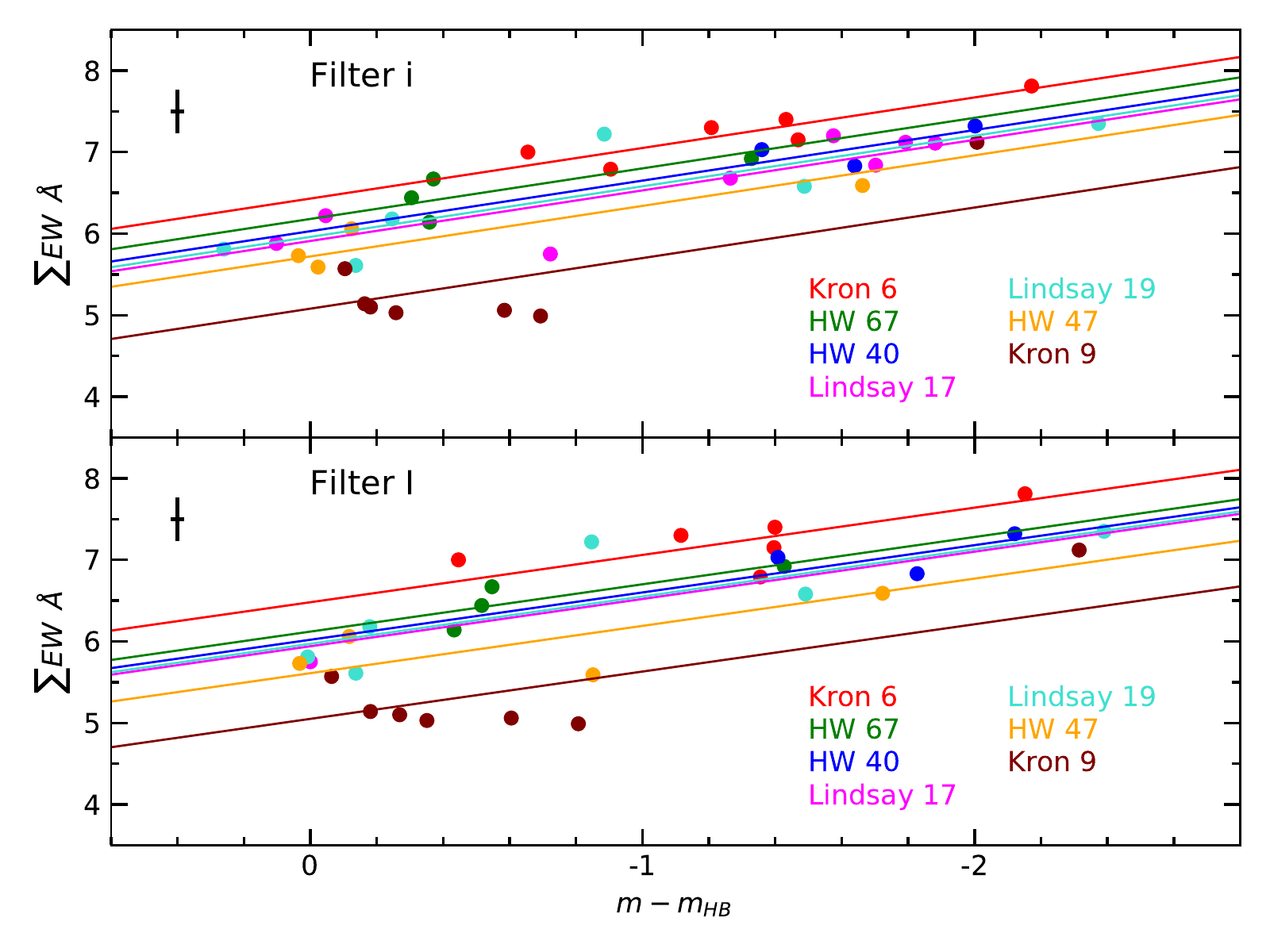}
      \caption{Same as Fig. \ref{fig:WV} for the $i$ and $I$ filters.}
         \label{fig:Wi}
   \end{figure}

\begin{figure}[!htb]
   \centering
   \includegraphics[width=0.9\columnwidth]{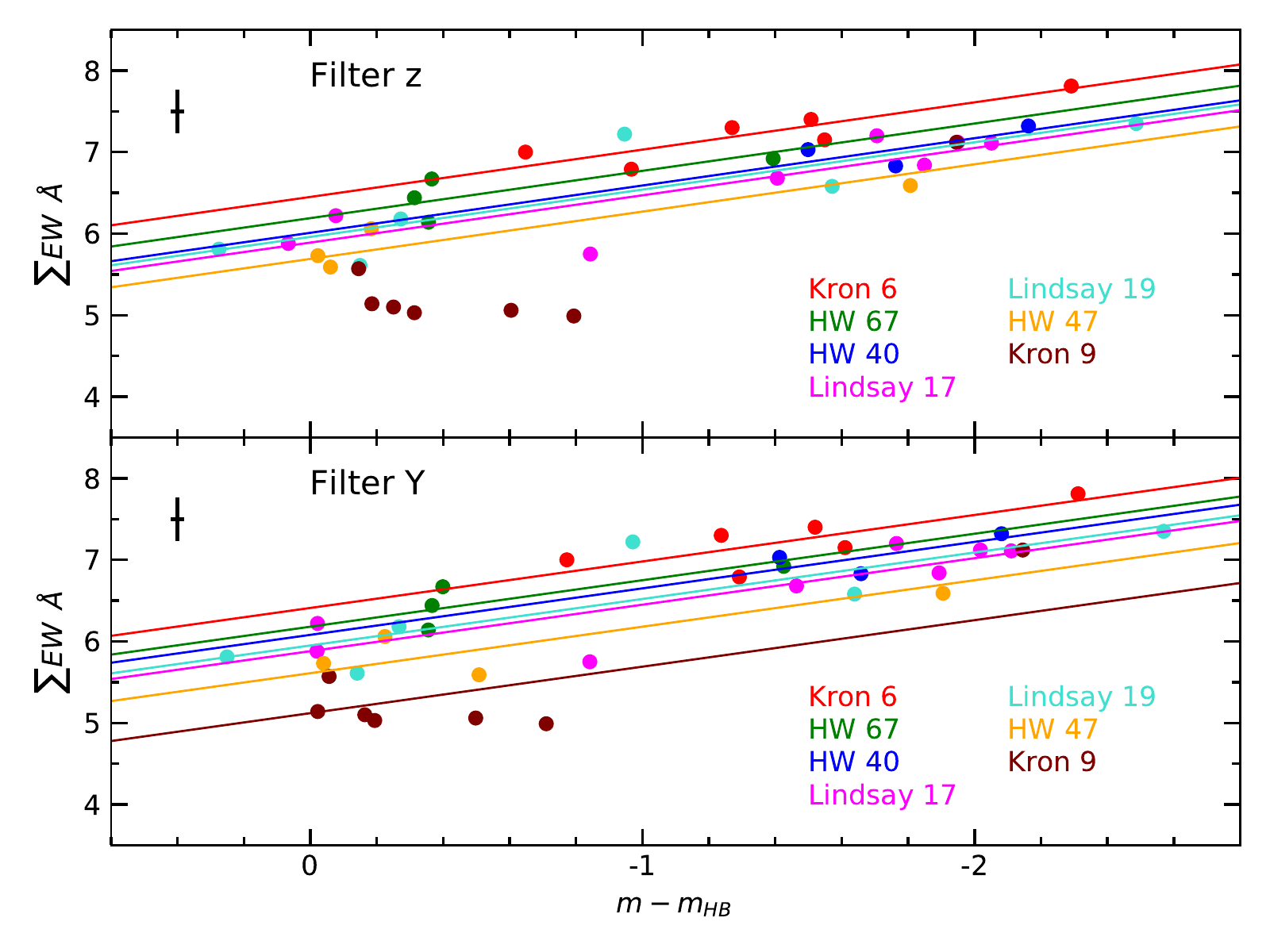}
      \caption{Same as Fig. \ref{fig:WV} for the $z$ and $Y$ filters.}
         \label{fig:Wz}
   \end{figure}

\begin{figure}[!htb]
   \centering
   \includegraphics[width=0.9\columnwidth]{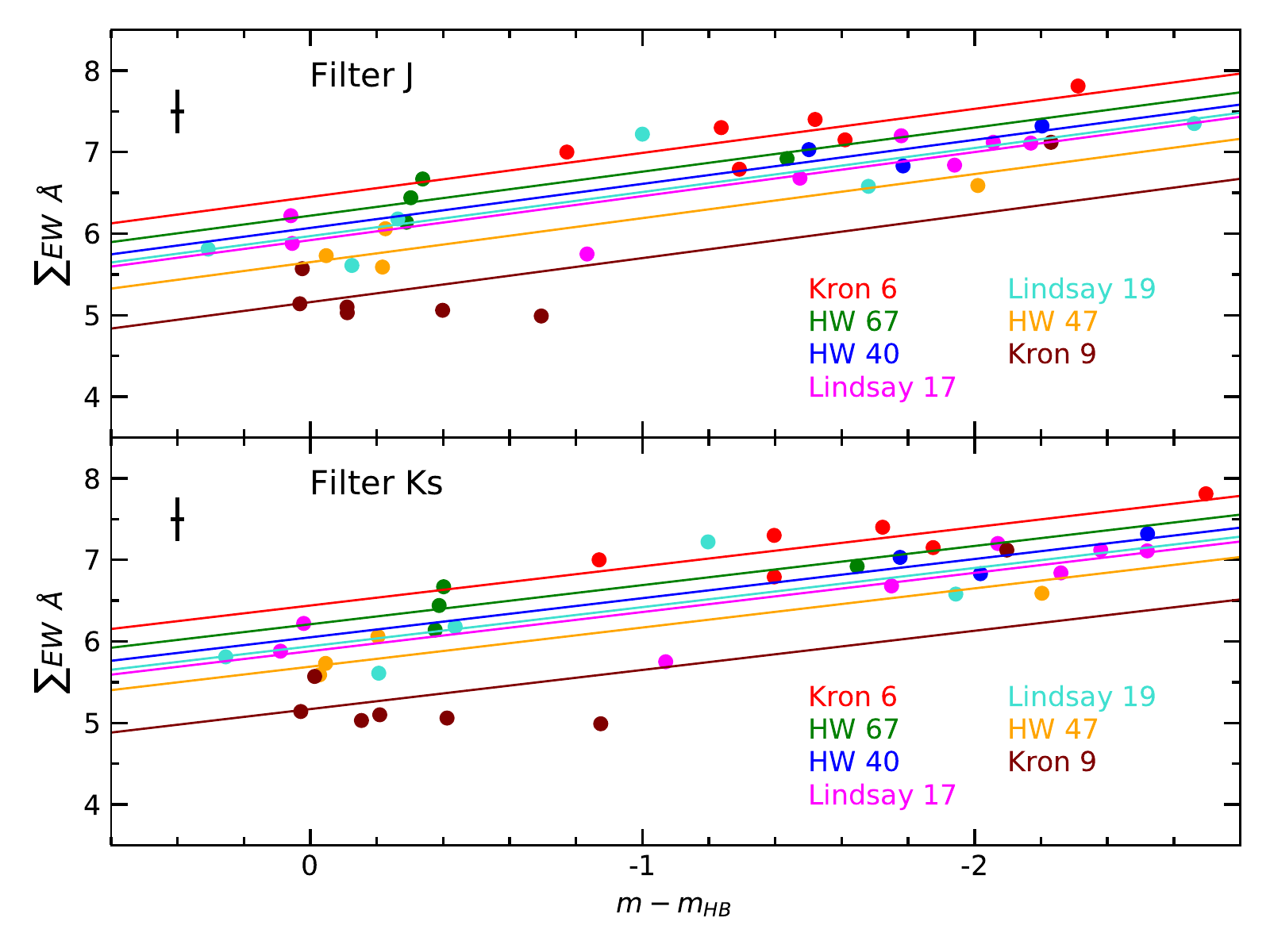}      
   \caption{Same as Fig. \ref{fig:WV} for the $J$ and $Ks$ filters.}
         \label{fig:WY}
   \end{figure}

In this section we show the Figs. \ref{fig:Wg} to \ref{fig:WY} for each filter, as showed in Fig. \ref{fig:WV}. The scales of the axes are all the same, thus it is possible to appreciate the slope change by eye from one filter to another.

\section{$W'_m$ - wavelength dependence}

\begin{figure}[!htb]
   \centering
   \includegraphics[width=0.9\columnwidth]{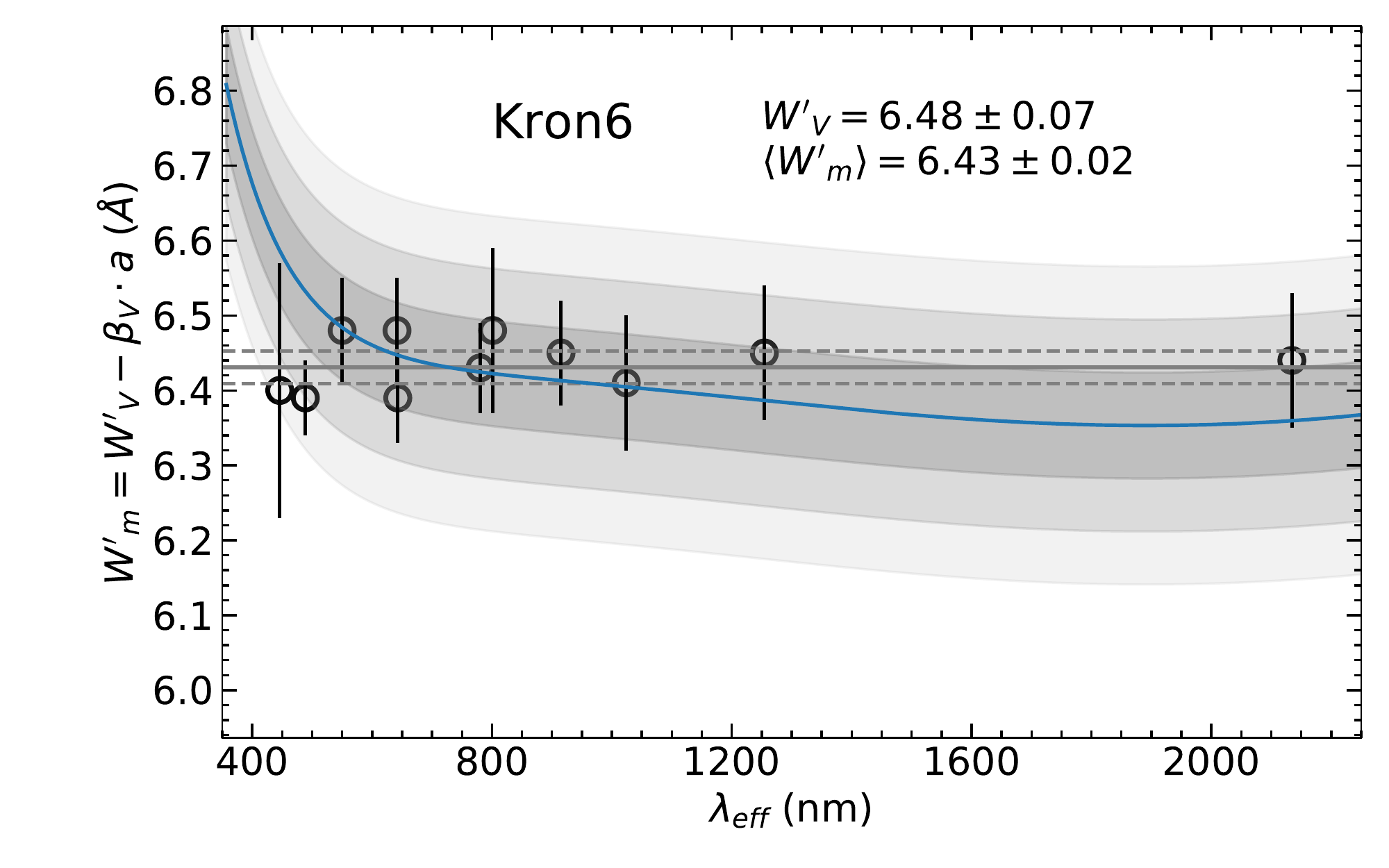}
      \caption{Same as Fig.\ref{fig:HW40Wall} for Kron\,6.}
         \label{fig:K6Wall}
   \end{figure}

\begin{figure}[!htb]
   \centering
   \includegraphics[width=0.9\columnwidth]{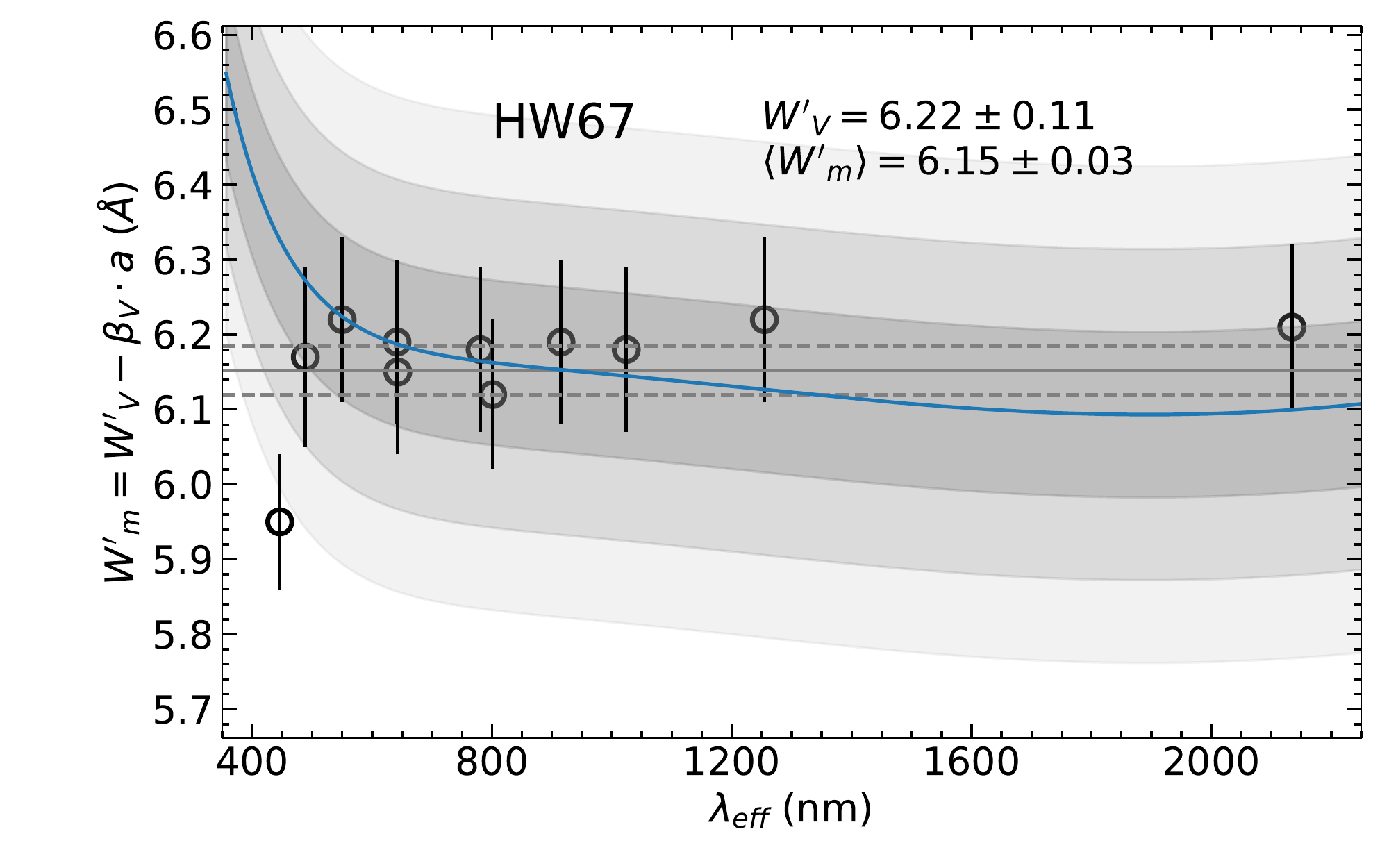}
      \caption{Same as Fig.\ref{fig:HW40Wall} for HW\,67.}
         \label{fig:HW67Wall}
   \end{figure}

\begin{figure}[!htb]
   \centering
   \includegraphics[width=0.9\columnwidth]{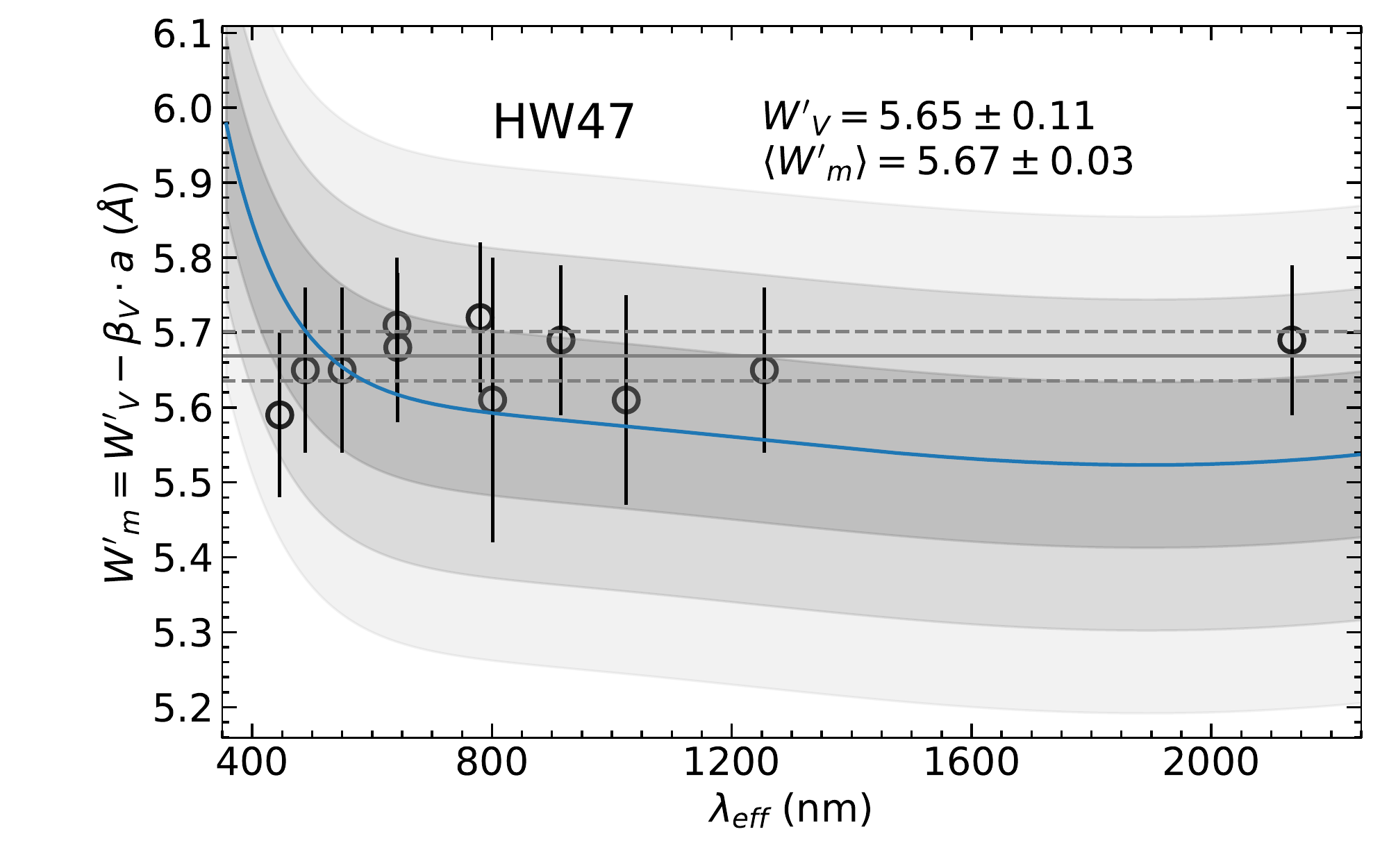}
      \caption{Same as Fig.\ref{fig:HW40Wall} for HW\,47.}
         \label{fig:HW47Wall}
   \end{figure}

\begin{figure}[!htb]
   \centering
   \includegraphics[width=0.9\columnwidth]{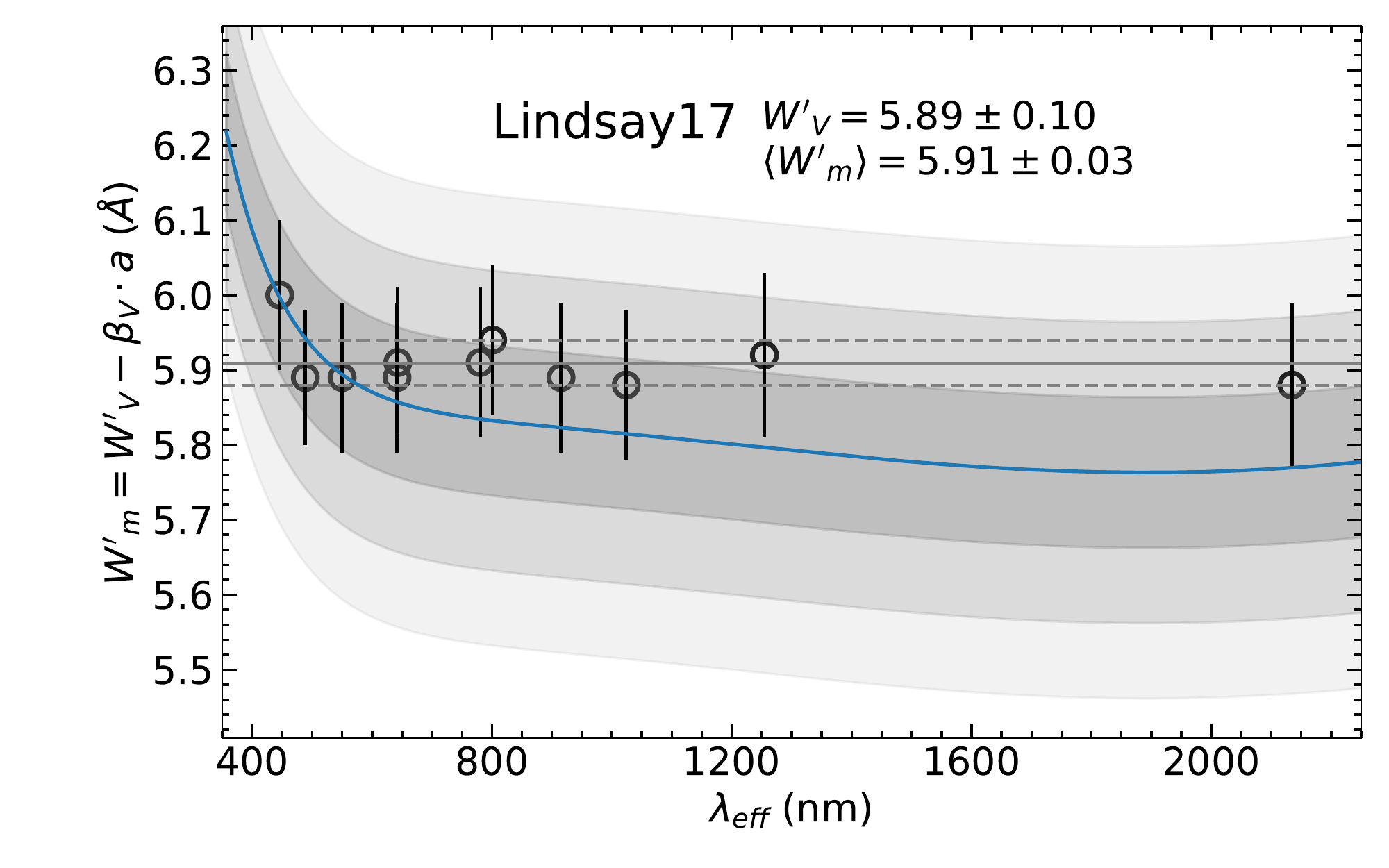}
      \caption{Same as Fig.\ref{fig:HW40Wall} for Lindsay\,17.}
         \label{fig:L17Wall}
   \end{figure}

\begin{figure}[!htb]
   \centering
   \includegraphics[width=0.9\columnwidth]{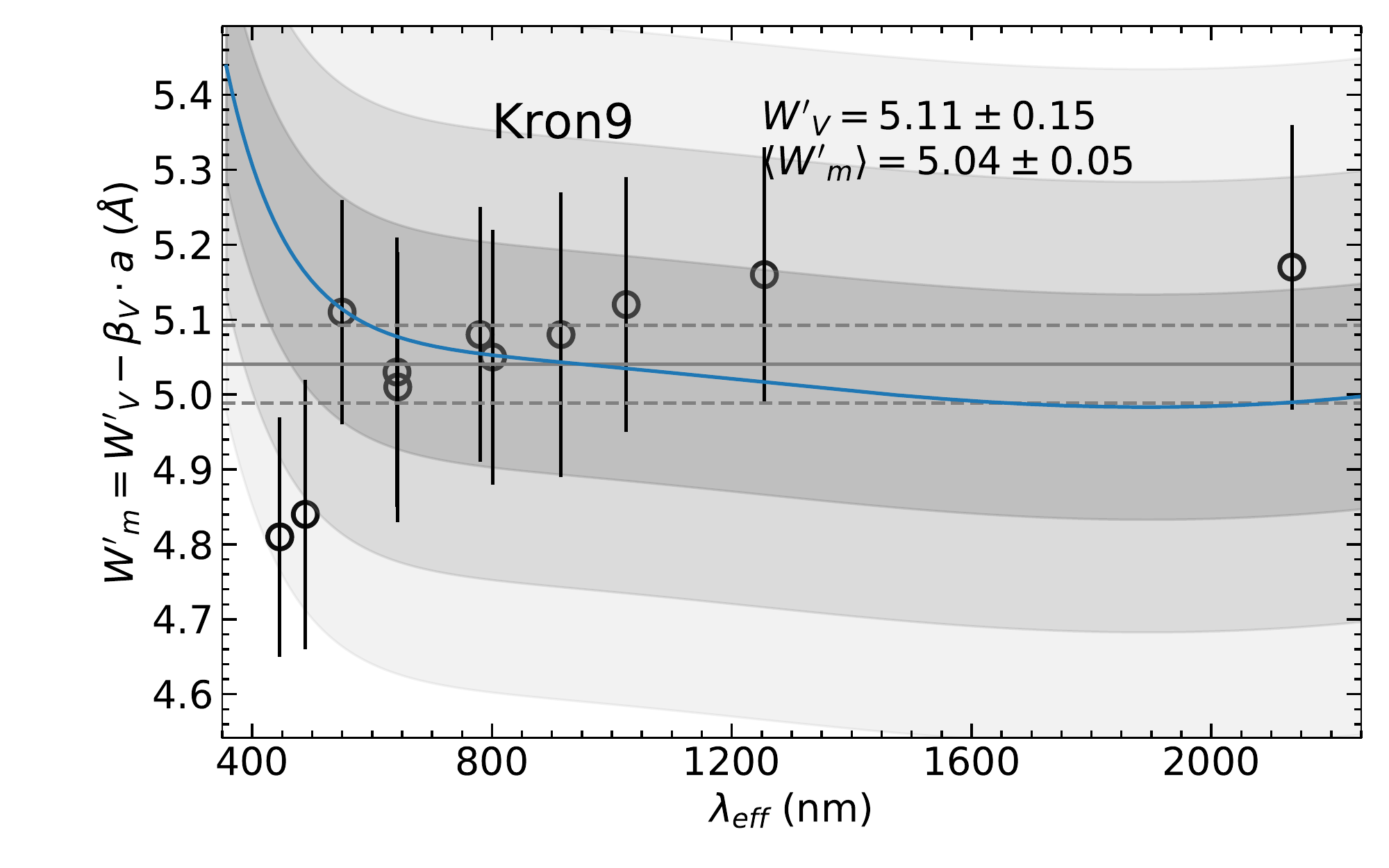}
      \caption{Same as Fig.\ref{fig:HW40Wall} for Kron\,9.}
         \label{fig:K9Wall}
   \end{figure}

\begin{figure}[!htb]
   \centering
   \includegraphics[width=0.9\columnwidth]{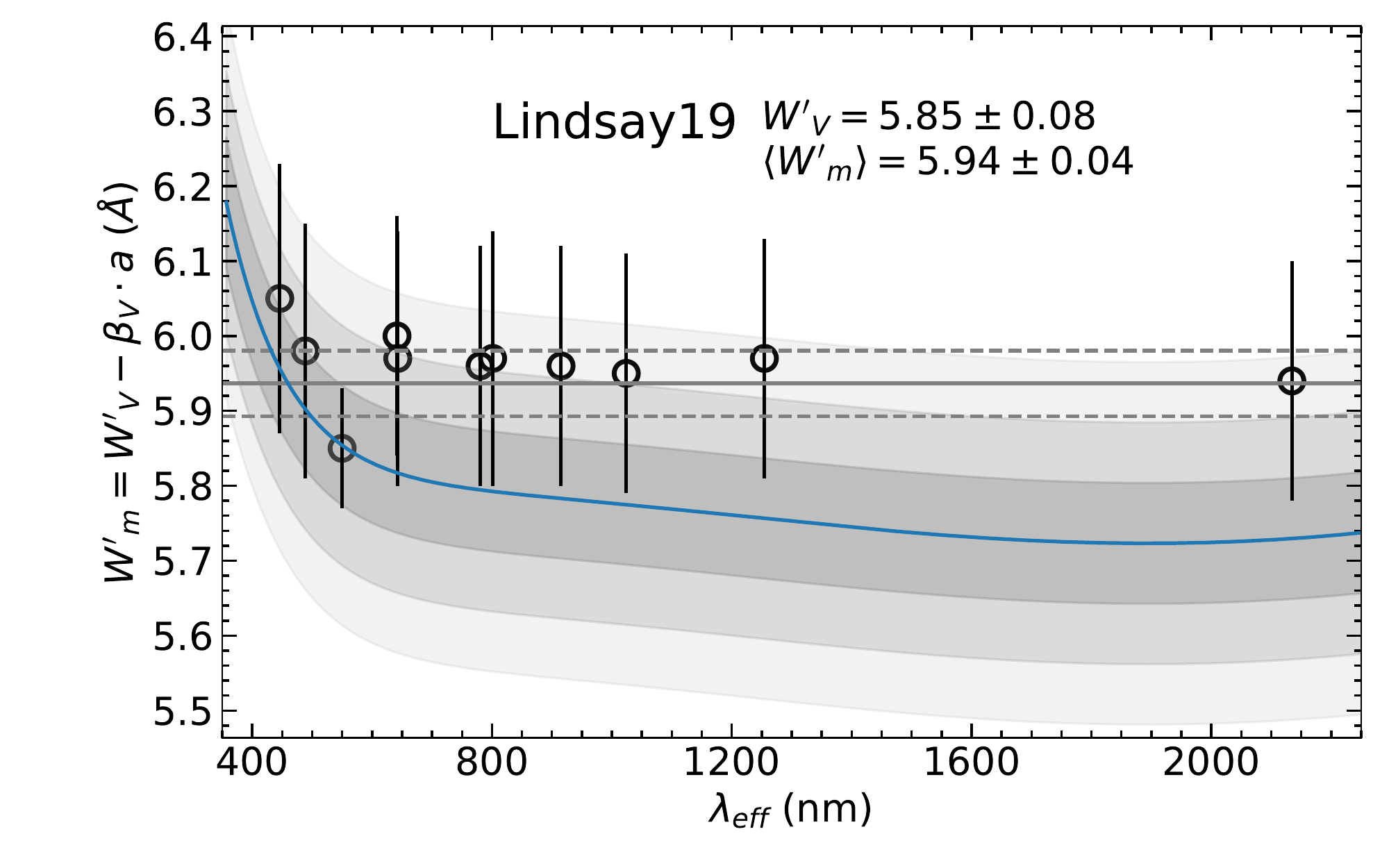}
      \caption{Same as Fig.\ref{fig:HW40Wall} for Lindsay\,19.}
         \label{fig:L19Wall}
   \end{figure}

In this section we show the Figs. \ref{fig:K6Wall} to \ref{fig:L19Wall} for each filter, as showed in Fig. \ref{fig:WV}. Some clusters show smaller dispersion than others, but overall, the fitted $W'_m$ seem to be fairly constant, at least for filters $V$ and redder. The fitted points are in agreement with the theoretical predictions within $1\sigma$, and no cluster follows the subtle predicted trend. An apparent border line case is Lindsay\,19, whose points are located $3\sigma$ away from the predicted function. The reason is because $W'_V$ is offset from the average of the rest of the points, and this is exactly the point used to anchor the relation.

\end{document}